\pdfoutput=1
\documentclass[11pt,twoside,a4paper,cmspaper,final,collab]{cms-tdr}
\def\svnVersion{3d2136f}\def\svnDate{2021/12/17}\def\cmsCernNoTag{CERN-EP-2021-126}\def\cmsCernDate{\today}\def\cmsMessage{Published in the Journal of High Energy Physics as \href{http://dx.doi.org/10.1007/JHEP12(2021)083}{\doi{10.1007/JHEP12(2021)083}.}}
\begin{document}\cmsNoteHeader{TOP-21-001}

\newcommand{\cmsTable}[1]{\resizebox{\textwidth}{!}{#1}}
\newlength\cmsTabSkip\setlength{\cmsTabSkip}{1ex}

\newcommand{\ctz}{\ensuremath{c_{\PQt \PZ}}\xspace}
\newcommand{\ctw}{\ensuremath{c_{\PQt \PW}}\xspace}
\newcommand{\cpqm}{\ensuremath{c^{-}_{\varphi \mathrm{Q}}}\xspace}
\newcommand{\cpqt}{\ensuremath{c^{3}_{\varphi \mathrm{Q}}}\xspace}
\newcommand{\cpt}{\ensuremath{c_{\varphi \PQt}}\xspace}
\newcommand{\otz}{\ensuremath{\mathcal{O}_{\PQt \PZ}}\xspace}
\newcommand{\otw}{\ensuremath{\mathcal{O}_{\PQt \PW}}\xspace}
\newcommand{\opqm}{\ensuremath{\mathcal{O}^{-}_{\varphi \mathrm{Q}}}\xspace}
\newcommand{\opqt}{\ensuremath{\mathcal{O}^{3}_{\varphi \mathrm{Q}}}\xspace}
\newcommand{\opt}{\ensuremath{\mathcal{O}_{\varphi \PQt}}\xspace}
\newcommand{\otg}{\ensuremath{\mathcal{O}_{\PQt \mathrm{G}}}\xspace}

\newcommand{\nnctztzq}{{NN-}{\ctz}{-}{\tZq}\xspace}
\newcommand{\nnctwtzq}{{NN-}{\ctw}{-}{\tZq}\xspace}
\newcommand{\nncpqttzq}{{NN-}{\cpqt}{-}{\tZq}\xspace}
\newcommand{\nncpqtttz}{{NN-}{\cpqt}{-}{\ttZ}\xspace}
\newcommand{\nnalltzq}{{NN-5D-}{\tZq}\xspace}
\newcommand{\nnctzttz}{{NN-}{\ctz}{-}{\ttZ}\xspace}
\newcommand{\nnctwttz}{{NN-}{\ctw}{-}{\ttZ}\xspace}
\newcommand{\nnallttz}{{NN-5D-}{\ttZ}\xspace}
\newcommand{\nnsm}{{NN-SM}\xspace}
\newcommand{\nneft}{{NN-EFT}\xspace}
\newcommand{\nnefts}{{NN-EFTs}\xspace}
\newcommand{\nnall}{{NN-5D}\xspace}

\newcommand*{\DeepJet}{\textsc{DeepJet}\xspace}
\newcommand*{\Keras}{\textsc{Keras}\xspace}
\newcommand*{\Tensorflow}{\textsc{TensorFlow}\xspace}
\newcommand*{\NNPDFThreeOne}{NNPDF3.1\xspace}
\newcommand*{\NNPDFThreeZero}{NNPDF3.0\xspace}
\newcommand*{\Adam}{\textsc{Adam}\xspace}
\newcommand*{\Madspin}{\textsc{Madspin}\xspace}

\newcommand{\pp}{{\Pp}{\Pp}\xspace}
\newcommand{\tZq}{{\PQt}{\PZ}{\PQq}\xspace}
\newcommand{\tWZ}{{\PQt}{\PW}{\PZ}\xspace}
\newcommand{\ttZ}{{\ttbar}{\PZ}\xspace}
\newcommand{\tttt}{{\ttbar}{\ttbar}\xspace}
\newcommand{\tHq}{{\PQt}{\PH}{\PQq}\xspace}
\newcommand{\tHW}{{\PQt}{\PH}{\PW}\xspace}
\newcommand{\ttg}{{\ttbar}{\PGg}\xspace}
\newcommand{\ttj}{{\ttbar}{\ensuremath{\text{+jets}}}\xspace}
\newcommand{\Wj}{{\PW}{\ensuremath{\text{+jets}}}\xspace}
\newcommand{\Zg}{{\PZ}{\PGg}\xspace}
\newcommand{\Vg}{{\PV}{\PGg}\xspace}
\newcommand{\ttH}{{\ttbar}{\PH}\xspace}
\newcommand{\ttHH}{{\ttbar}{\PH}{\PH}\xspace}
\newcommand{\ttW}{{\ttbar}{\PW}\xspace}
\newcommand{\ttVH}{{\ttbar}{\PV}{\PH}\xspace}
\newcommand{\ttVV}{{\ttbar}{\PV}{\PV}\xspace}
\newcommand{\dy}{\ensuremath{{\PZ}/{\PGg^\star}\xspace}}
\newcommand{\ZZ}{{\PZ}{\PZ}\xspace}
\newcommand{\WZ}{{\PW}{\PZ}\xspace}
\newcommand{\WWW}{{\PW}{\PW}{\PW}\xspace}
\newcommand{\WWZ}{{\PW}{\PW}{\PZ}\xspace}
\newcommand{\WZZ}{{\PW}{\PZ}{\PZ}\xspace}
\newcommand{\ZZZ}{{\PZ}{\PZ}{\PZ}\xspace}
\newcommand{\VVV}{{\PV}{\PV}{(\PV)}\xspace}
\newcommand{\Xg}{{X}{\PGg}\xspace}
\newcommand{\tX}{{\PQt}{(\PAQt)}{X}\xspace}

\newcommand{\zpt}{\ensuremath{\pt^{\PZ}}\xspace}
\newcommand{\costhetaz}{\ensuremath{\cos\theta^{\star}_{\PZ}}\xspace}
\newcommand{\mtw}{\ensuremath{\mT^{\PW}}\xspace}
\newcommand{\abseta}{\ensuremath{\abs{\eta}}\xspace}
\newcommand{\mz}{\ensuremath{m_\PZ}\xspace}
\newcommand{\mlll}{\ensuremath{m_{3\ell}}\xspace}

\newcommand{\srtzq}{\ensuremath{\text{SR-}{\tZq}}\xspace}
\newcommand{\srttz}{\ensuremath{\text{SR-}{\ttZ}}\xspace}
\newcommand{\srother}{\ensuremath{\text{SR-Others}}\xspace}
\newcommand{\srttzfour}{\ensuremath{\text{SR-}{\ttZ}\text{-}4\ell}\xspace}
\newcommand{\sr}{\ensuremath{\text{SR-}3\ell}\xspace}
\newcommand{\usedLumi}{\ensuremath{138\fbinv}\xspace}
\newcommand{\sW}{\ensuremath{s_{\mathrm{W}}}\xspace}
\newcommand{\cW}{\ensuremath{c_{\mathrm{W}}}\xspace}

\newcommand*{\rot}[1]{\rotatebox{90}{#1}}

\cmsNoteHeader{TOP-21-001}
\title{Probing effective field theory operators in the associated production of top quarks with a \texorpdfstring{\PZ}{Z} boson in multilepton final states at \texorpdfstring{$\sqrt{s} = 13\TeV$}{sqrt(s) = 13 TeV}}
\date{\today}

\abstract{
A search for new top quark interactions is performed within the framework of an effective field theory using the associated production of either one or two top quarks with a \PZ boson in multilepton final states.  The data sample corresponds to an integrated luminosity of \usedLumi of proton-proton collisions at $\sqrt{s} = 13\TeV$ collected by the CMS experiment at the LHC.  Five dimension-six operators modifying the electroweak interactions of the top quark are considered.  Novel machine-learning techniques are used to enhance the sensitivity to effects arising from these operators.  Distributions used for the signal extraction are parameterized in terms of Wilson coefficients describing the interaction strengths of the operators.  All five Wilson coefficients are simultaneously fit to data and 95\% confidence level intervals are computed.  All results are consistent with the SM expectations.
\\[1cm]
\begin{flushright}
\textit{We dedicate this publication to our friend and colleague Nicolas Tonon\\ who passed away unexpectedly while this paper was in print.\\ This work would not have been possible without Nicolas' devotion and major contributions.}
\end{flushright}
}

\hypersetup{
pdfauthor={CMS Collaboration},
pdftitle={Probing effective field theory operators in the associated production of top quarks with a Z boson in multilepton final states at sqrt(s)=13 TeV},
pdfsubject={CMS},
pdfkeywords={CMS, EFT, top quark, Z boson, machine learning, multilepton, MVA}}
\maketitle

\section{Introduction}
\label{sec:intro}
The standard model (SM) of particle physics is supported by a vast number of measurements from experiments covering a broad energy range up to the TeV scale.
However, the SM falls short of explaining several observed phenomena, such as neutrino oscillations~\cite{PDG2020} and the baryon asymmetry in our universe~\cite{Canetti_2012}, and cannot accommodate the existence of dark matter and dark energy~\cite{ARUN2017166}. 
More generally, there are many indications that the SM corresponds to a low-energy approximation of a more fundamental theory beyond the standard model (BSM). 
Several BSM models explain these phenomena by postulating the existence of new particles or mechanisms. 
However, these hypothetical new particles may be too massive to be directly accessible at the CERN LHC.

Even in the absence of any direct observation of a new particle, new phenomena may manifest themselves indirectly through corrections at the quantum-loop level, thus leading to observable deviations in well-established processes.
Such deviations can be interpreted in a coherent and model-independent manner using the approach of effective field theory (EFT)~\cite{BUCHMULLER1986621,eft_intro}. 
An EFT corresponds to an approximation at low energy of an underlying theory characterized by an energy scale $\Lambda$ that is well above the typical energies accessible at colliders.
New, effective interactions between the SM fields are introduced by extending the SM Lagrangian with higher-order operators. 
The interaction strength of an operator of dimension $d$ is characterized by a dimensionless Wilson coefficient (WC) and is proportional to $1/\Lambda^{d-4}$. 
This factor suppresses the contribution from higher-order operators, implying that effects of new interactions can be approximated with a finite set of WCs associated with lower-order operators.
Since operators with $d=5$ or 7 violate lepton and/or baryon number conservation~\cite{HELSET2020135132}, we restrict our analysis to dimension-six operators.
The effective Lagrangian can then be written as
\begin{linenomath}
\begin{equation}
 \mathcal{L}_{\text{eff}} = \mathcal{L}_{\text{SM}} + \sum_{i} \frac{c_{i}}{\Lambda^{2}} \mathcal{O}_{i} ,
\end{equation}
\end{linenomath}
where $\mathcal{L}_{\text{SM}}$ is the SM Lagrangian, $\mathcal{O}_{i}$ are dimension-six operators, and $c_{i}$ are the corresponding WCs that can be constrained from experimental data. 

\begin{figure}[!hbtp]
\centering \includegraphics[width=0.46\textwidth]{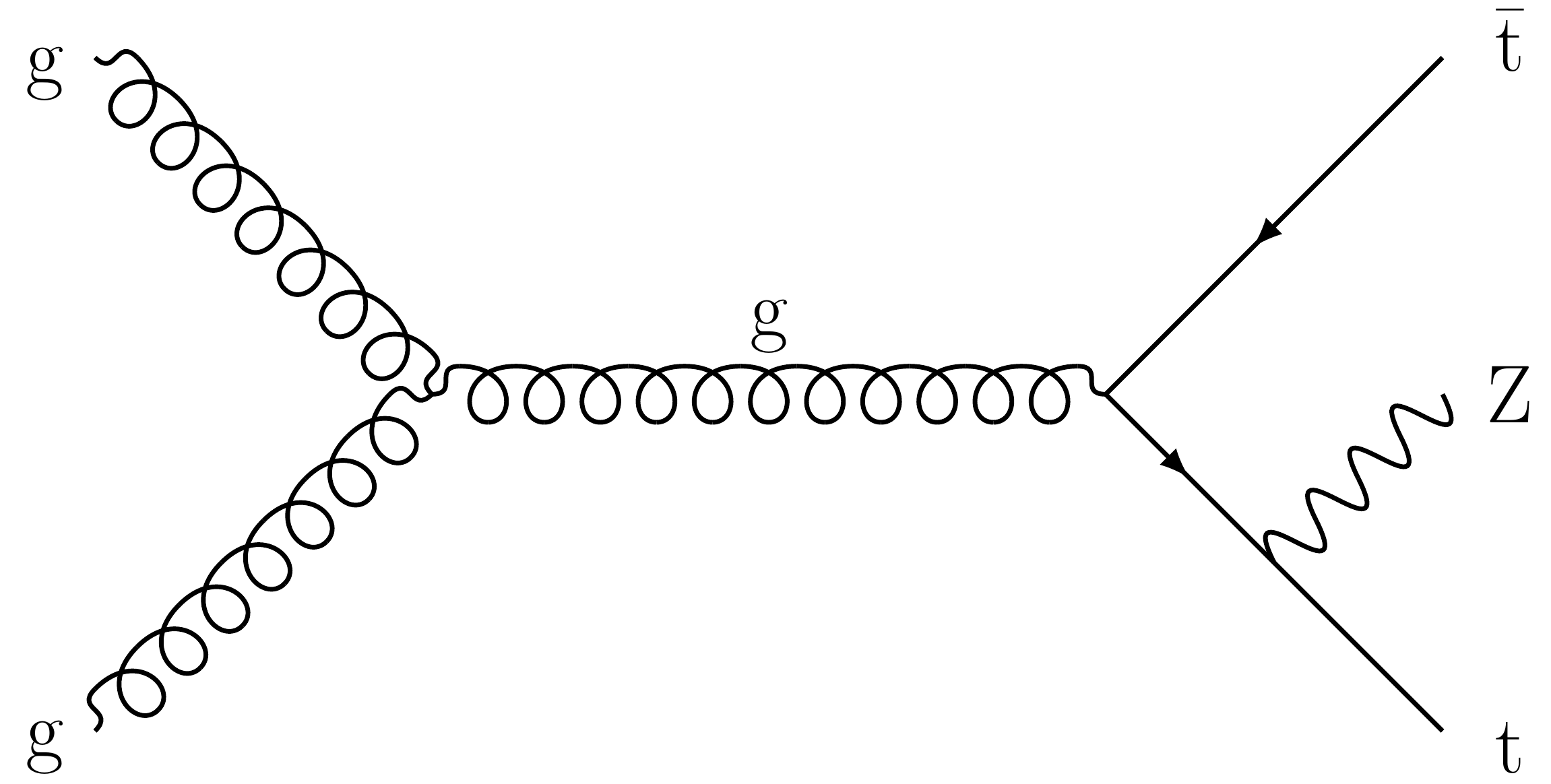} \hfill
\centering \includegraphics[width=0.44\textwidth]{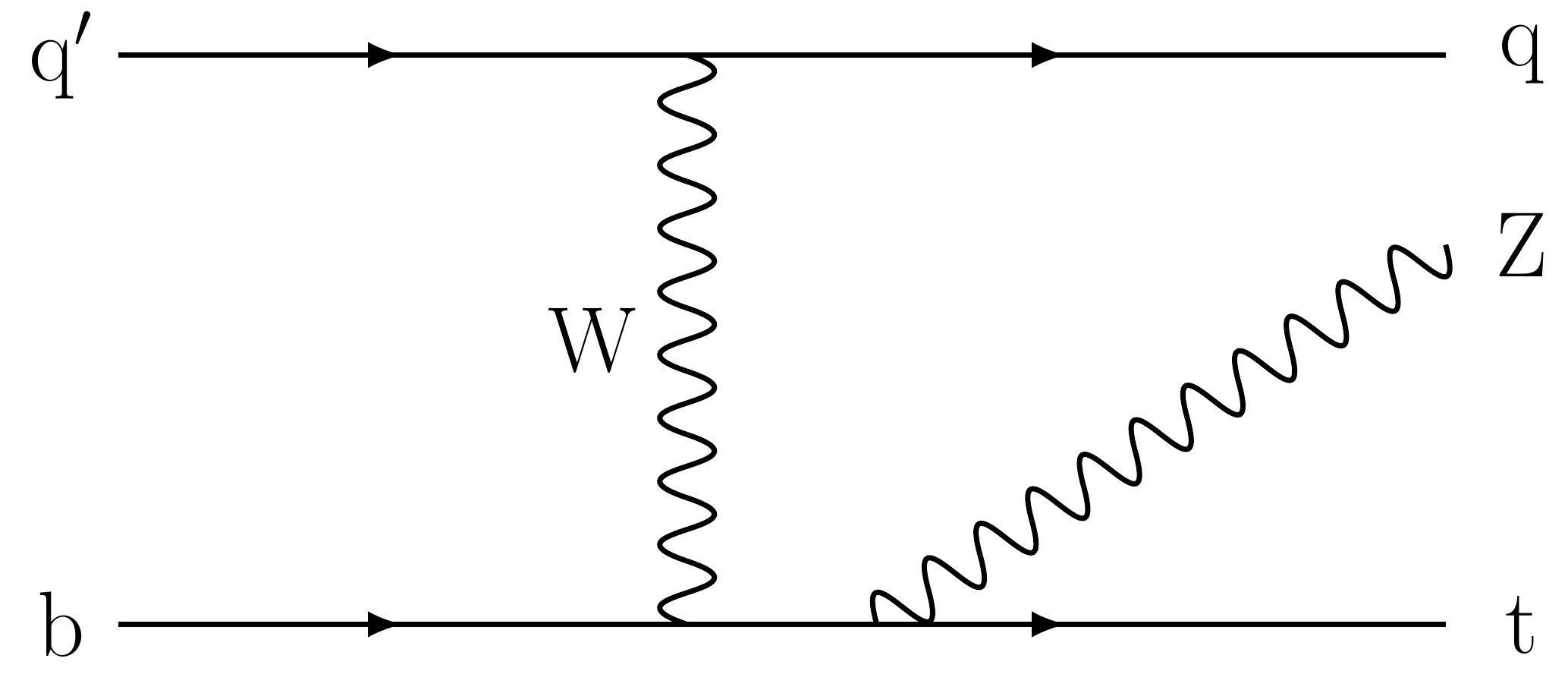}
\centering \includegraphics[width=0.47\textwidth]{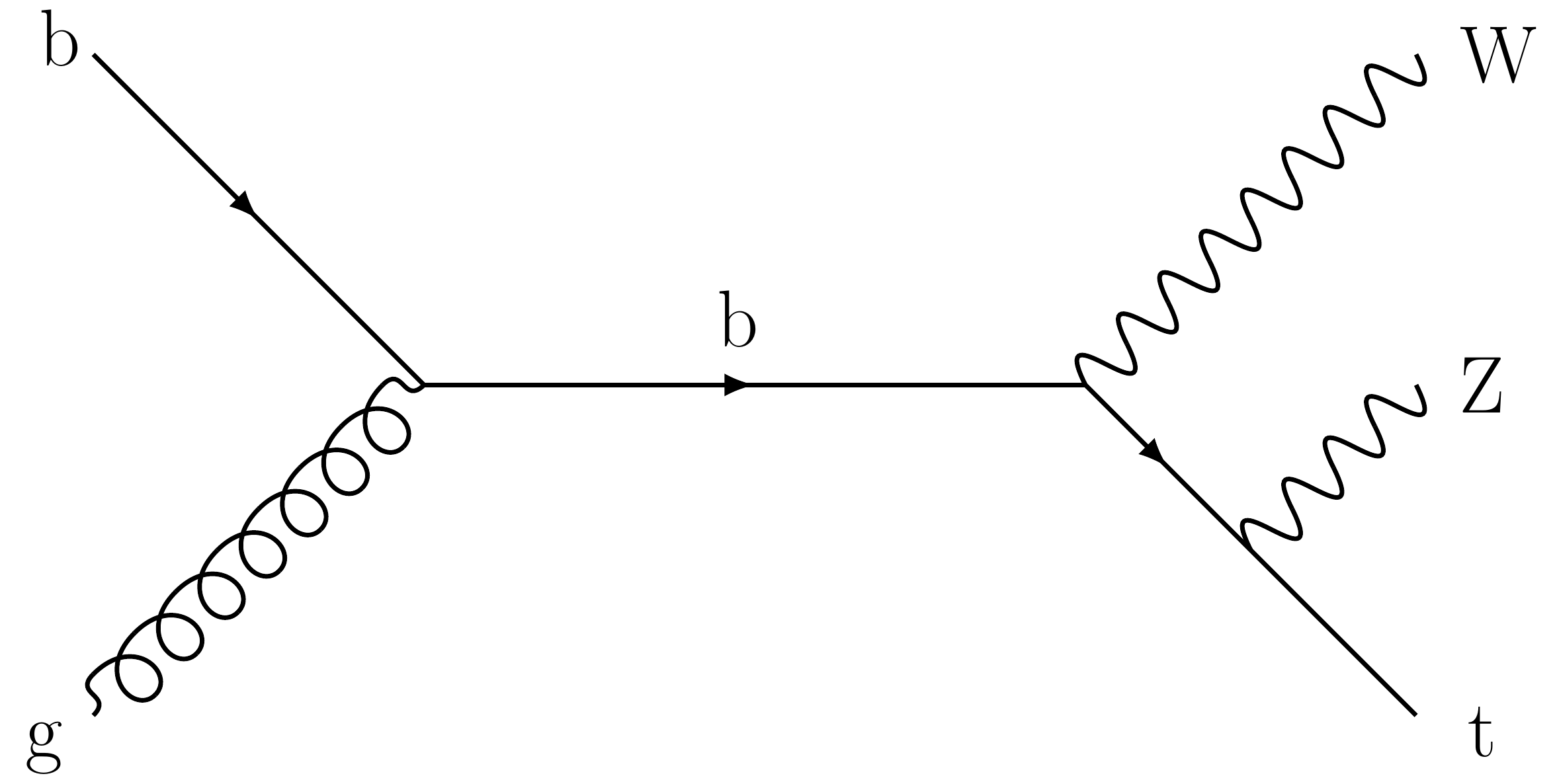}
\caption{Representative Feynman diagrams at tree level for \ttZ (upper left), \tZq (upper right), and \tWZ (lower) production.}
\label{fig:feynman}
\end{figure}

The large top quark mass $m_{\PQt} = 172.44 \pm 0.49\GeV$~\cite{topMass_CMS} corresponds to a Yukawa coupling to the Higgs boson close to unity. 
This suggests that the top quark may play a special role within the SM, and that its precise characterization may shed light on the electroweak symmetry breaking mechanism~\cite{Dobrescu:1997nm,Chivukula:1998wd,Delepine:1995qs}. 
The high integrated luminosity and center-of-mass energy at the LHC make it possible to study rare top quark processes, such as the associated production of top quarks with a \PZ\ boson, where top quarks are either produced in pairs (\ttZ), or singly in the \tZq and \tWZ channels. 
Representative tree-level Feynman diagrams are shown for the three signal processes in Fig.~\ref{fig:feynman}. 
Whereas the \ttZ and \tZq processes have comparable inclusive cross sections of about 800\unit{fb} and were already observed by both the ATLAS and CMS Collaborations~\cite{Sirunyan:2018zgs,tzq_atlas_obs,Khachatryan:2015sha,CERN-EP-2021-011}, the \tWZ process has a much smaller cross section and has not been observed yet.
These three processes are of major interest because they probe the coupling of the top quark to the \PZ\ boson at tree level. 
Numerous BSM extensions predict sizable modifications to this coupling~\cite{Grojean2013,PhysRevD.84.015003,Chivukula:1994mn}, 
which is among the least constrained by the available data in the top quark sector.

The focus of the present analysis is the study of the electroweak interactions of the top quark, and in particular the \ttZ interaction.
Consequently, we only consider EFT operators involving third-generation quarks and gauge bosons, which interfere with the SM production of the \ttZ or \tZq signal. 
This means that these operators must affect either process---or both---at order $1/\Lambda^{2}$. 
We restrict our study to CP-conserving effects, and thus we ignore the imaginary components of complex WCs. 
Finally, the \otg operator related to the chromomagnetic dipole moment of the top quark is ignored, since it can be probed with much better sensitivity in \ttbar events~\cite{PhysRevD.100.072002}.
As a result, we focus on a subset of five operators, namely: \otz, \otw, \opqt, \opqm, and \opt~\cite{lhctopwg}. 
The \otz and \otw operators induce electroweak dipole moments of the top quark, \opqt is the left-handed $\mathrm{SU}(2)$ triplet current operator, and the \opqm and \opt neutral-current operators modify the interactions of the \PZ boson with left- and right-handed top quarks, respectively. 
Comprehensive descriptions of their effects on top quark interactions are given in Refs.~\cite{Brivio:2019ius,Maltoni:2019aot}.

A key aspect of this study is the use of novel multivariate analysis (MVA) techniques based on machine learning to enhance the sensitivity to new phenomena arising from the EFT operators.
Since new operators usually affect the distributions of multiple observables, MVA techniques that exploit correlations in high-dimensional data are well suited for EFT measurements. 
We train machine-learning algorithms for two purposes.
Firstly, a multiclass classifier is trained to distinguish between different SM processes, and is used to define subregions enriched either in signal or background events. 
Secondly, binary classifiers are trained to separate events generated according to the SM from events generated with nonzero WC values for one or more EFT operators. 
They are used to construct powerful discriminating observables that are ultimately fit to data to compute two confidence intervals for each WC, one keeping the other WCs fixed to zero and the other treating all five WCs as free parameters. 
The core ideas for these binary classifiers appeared recently in the literature and since then have garnered increased attention, notably because they were shown to outperform traditional approaches based on single observables in several case studies at the generator level~\cite{eft_ml1,hollingsworth2020resonance,PhysRevD.100.035040}. 
This motivates the application of this technique for the first time in an LHC analysis involving the interference between EFT operators and the SM amplitude.

The binary classifiers are trained with simulated samples whose event weights are parameterized as functions of the five WCs of interest. 
These samples are passed through a full detector simulation, and are used to search for new interactions without making any simplifying assumption regarding the parton shower and detector response.
A previous CMS analysis in the top quark sector employed this approach to parameterize new interactions directly at the detector level in the context of an EFT~\cite{TOP-19-001}. It used data collected in 2017 and targeted multilepton final states.
That analysis set constraints on 16 WCs simultaneously by performing counting experiments in various event categories, which were defined based on the jet and lepton multiplicities, the presence of a \PZ\ boson candidate, and the sum of the lepton charges.

The data sample of proton-proton (\pp) collisions at $\sqrt{s}=13\TeV$ used in this paper was collected by the CMS experiment during Run~2 of the LHC (2016--2018) and corresponds to an integrated luminosity of \usedLumi, of which 36.3~\cite{CMS-LUM-17-003}, 41.5~\cite{LUM-17-004}, and 59.7~\cite{LUM-18-002}\fbinv were recorded in 2016, 2017, and 2018, respectively.
We target events in which the decays of the top quark and \PZ\ boson lead to final states with three or four leptons, either electrons or muons; 
this also includes a small contribution from leptonic tau lepton decays.
Unless stated otherwise and throughout this paper, the term lepton refers exclusively to electrons and muons, generically denoted by $\ell$.
These multilepton final states offer high trigger efficiencies, and a more favorable signal-to-background ratio compared with hadronic channels. 

The paper is organized as follows. 
We provide a brief overview of the CMS detector in Section~\ref{sec:detector}. 
Section~\ref{sec:samples} describes the data and simulated event samples, as well as the parameterization of the weights for simulated events.
Section~\ref{sec:reconstruction} details the object and event reconstructions, 
while the event selection and categorization are presented in Section~\ref{sec:selection}.
The estimation of backgrounds is presented in Section~\ref{sec:backgrounds}. 
The MVA is described in Section~\ref{sec:mva}, the systematic uncertainties affecting the measurements in Section~\ref{sec:systematics}, and the signal extraction procedure in Section~\ref{sec:extraction}. 
We discuss the results in Section~\ref{sec:results}, and conclude with a summary of the paper in Section~\ref{sec:summary}. 
Tabulated results are provided in HEPData~\cite{hepdata}.

\section{The CMS detector}
\label{sec:detector}
The central feature of the CMS apparatus is a superconducting solenoid of 6\unit{m} internal diameter, providing a magnetic field of 3.8\unit{T}. Within the solenoid volume are a silicon pixel and strip tracker, a lead tungstate crystal electromagnetic calorimeter (ECAL), and a brass and scintillator hadron calorimeter, each composed of a barrel and two endcap sections. Forward calorimeters extend the pseudorapidity ($\eta$) coverage provided by the barrel and endcap detectors. Muons are detected in gas-ionization chambers embedded in the steel flux-return yoke outside the solenoid. 

Events of interest are selected using a two-tiered trigger system. The first level (L1), composed of custom hardware processors, uses information from the calorimeters and muon detectors to select events at a rate of around 100\unit{kHz} within a fixed latency of about 4\mus~\cite{Sirunyan:2020zal}. The second level, known as the high-level trigger, consists of a farm of processors running a version of the full event reconstruction software optimized for fast processing, and reduces the event rate to around 1\unit{kHz} before data storage~\cite{Khachatryan:2016bia}. 
A more detailed description of the CMS detector, together with a definition of the coordinate system and of the relevant kinematic variables, can be found in Ref.~\cite{Chatrchyan:2008zzk}.

\section{Data sample and Monte Carlo simulations}
\label{sec:samples}
The data sample was recorded using a combination of single-, double-, and triple-lepton trigger algorithms, 
whose thresholds on the transverse momentum \pt with respect to the beam axis vary between data-taking periods depending on the instantaneous luminosity. 
For instance, the minimum \pt threshold of the single-electron (-muon) triggers ranges between 25--35 (22--27)\GeV. 
For events selected in this analysis, the combined trigger efficiency is nearly 100\% both in data and simulation.

Event samples produced via Monte Carlo (MC) simulation are used to estimate the contributions of signal processes and most background processes, 
as well as to train machine-learning algorithms.
The signal samples are generated at leading order (LO) in perturbative quantum chromodynamics (QCD) and incorporate EFT effects, whereas background samples do not include EFT effects and are generated at next-to-LO (NLO) whenever possible.
Additional signal samples generated at NLO without EFT effects are used for validation purposes, 
and to train a classification algorithm aiming to separate different SM processes, as described in Section~\ref{sec:mva}. 
The latter samples will be referred to explicitly as ``SM signal samples''. The mass of the top quark is set to $m_{\PQt} = 172.5 \GeV$ in simulation.

The SM signal samples for the \tZq and \ttZ processes, as well as the samples for several background processes (\WZ, \ttW, \tttt, multiboson production, and \Vg, where $\text{V}$ denotes either a \PW or \PZ boson), are generated at NLO using $\MGvATNLO$ 2.4.2~\cite{Alwall_2014}. 
The SM \tWZ sample and other background samples are generated at LO using $\MGvATNLO$ 2.4.2 (\tHq, \tHW, \ttg, \ttVV, \ttVH, \ttHH) or \MCFM 7.0.1~\cite{MCFM1, MCFM2} 
($\Pg\Pg \to$ \ZZ), or at NLO using \POWHEG 2.0~\cite{Nason:2004rx, Frixione:2007vw, Alioli:2010xd} ($\qqbar \to \ZZ$~\cite{Nason:2013ydw}, \ttH~\cite{Hartanto:2015uka}).
The SM \tZq sample is generated in the four-flavor scheme, in which only up, down, strange, and charm quarks are considered as sea quarks of the proton, whereas the SM \tWZ sample is generated in the five-flavor scheme, in which bottom quarks are considered as sea quarks of the proton as well and may appear in the initial state of \pp scattering processes~\cite{Maltoni:2012pa}. 

The \NNPDFThreeOne~\cite{NNPDF31} set of parton distribution functions (PDFs) is used to simulate signal samples for all three years, and background samples for 2017--2018, and the \NNPDFThreeZero~\cite{NNPDF30} set of PDFs is used to generate background samples for 2016.
The parton showering, hadronization, and underlying event are modeled using \PYTHIA 8.2~\cite{Sjostrand:2014zea} with the tune CP5~\cite{Sirunyan:2019dfx} for the 2017--2018 samples, as well as for the 2016 samples for the signal, \ttW, \ttH, \ttg, \Zg, and \tttt processes, whereas the tunes CUETP8M1 or CUETP8M2T4~\cite{CMS-PAS-TOP-16-021,Khachatryan:2015pea} are used to simulate other background samples for 2016. 
The matching of matrix elements (MEs) to the parton shower (PS) is done using the FxFx~\cite{Frederix2012} merging scheme for NLO samples, and the MLM scheme~\cite{Alwall_2007} for LO samples.

The presence of simultaneous \pp collisions in the same or nearby bunch crossings, referred to as pileup, is modeled by superimposing inelastic \pp interactions simulated using \PYTHIA 8.2 on all generated events. 
Generated events are passed through a detailed simulation of the CMS detector based on \GEANTfour~\cite{GEANT4}, and are reconstructed with the same version of the CMS event reconstruction software used for data.

Residual differences between data and simulation are corrected by modifying the weights of generated events, or by varying relevant simulated quantities.
Such differences are observed in: the pileup distribution; the reconstruction and identification efficiencies for electrons and muons; the jet energy scale and resolution; the efficiency to identify jets originating from the hadronization of bottom quarks and the corresponding misidentification rates for light (\PQu, \PQd, \PQc, \PQs)-quark and gluon jets; and the resolution in missing transverse momentum. 

\subsection{Simulation of the signal samples}
\label{sec:signal}
The signal samples including EFT effects are generated at LO using $\MGvATNLO$ 2.6.5 and the \NNPDFThreeOne PDF set. 
The decays of top quarks and \PW bosons are simulated with the \Madspin program~\cite{Artoisenet:2012st}. 

We generate the signal events following a similar approach to that outlined in Ref.~\cite{lhctopwg}.
The EFT model used in the present analysis focuses on dimension-six operators that give rise to interactions involving at least one top quark. 
The degrees of freedom of this model are defined as linear combinations of Warsaw-basis operator coefficients~\cite{Grzadkowski:2010es}, and the mapping between both bases is given in Table~\ref{tab:operators}.
Since this model only allows for tree-level generation, the \ttZ sample is generated with an extra parton in the final state to improve its accuracy. 
The MLM merging scheme is used to match the MEs to the PS\@. 
However, the \tZq sample does not include an extra final-state parton because this matching procedure cannot be performed correctly for single top quark processes in the $t$ channel (due to the presence of a final-state light quark in the MEs), and neither does the \tWZ sample to avoid overlap with \ttZ production. 
We consider Feynman diagrams including at most one EFT vertex, in which the top quark is produced.

\begin{table}[!hbtp]
\centering
\topcaption{
  \label{tab:operators}
  List of dimension-six EFT operators considered in this analysis and their corresponding WCs. 
  The linear combinations of WCs to which they correspond in the Warsaw basis are indicated. \
  The abbreviations $\sW$ and $\cW$ denote the sine and cosine of the weak mixing angle, respectively. 
  The definitions of the relevant Warsaw-basis operators can be found in Ref.~\cite{lhctopwg}.}
\begin{tabular}{lll}
    Operator & WC & Mapping to Warsaw-basis coefficients \\ 
    \hline \\[-2ex]
    \otz    & \ctz  & $\text{Re} \big\{ -\sW c^{(33)}_{\mathrm{uB}} + \cW c^{(33)}_{\mathrm{uW}} \big\}$ \\ [\cmsTabSkip]
    \otw    & \ctw  & $\text{Re} \big\{ c^{(33)}_{\mathrm{uW}} \big\}$ \\ [\cmsTabSkip]
    \opqt   & \cpqt & $c^{3(33)}_{\varphi \mathrm{q}}$ \\ [\cmsTabSkip]
    \opqm   & \cpqm & $c^{1(33)}_{\varphi \mathrm{q}} - c^{3(33)}_{\varphi \mathrm{q}}$ \\ [\cmsTabSkip]
    \opt    & \cpt  & $c^{(33)}_{\varphi \mathrm{u}}$ \\ [\cmsTabSkip]
\end{tabular}
\end{table}

\subsection{Parameterization of the event weights} 
\label{sec:parameterization}
To interpret potential deviations in signal production in terms of new interactions, we parameterize the weights of the simulated signal events with the five WCs of interest.
Within an EFT framework, a given ME may be decomposed into its SM and EFT components as
\begin{linenomath} 
\begin{equation}
 \mathcal{M} = \mathcal{M}_{\text{SM}} + \mathcal{M}_{\text{EFT}} = \mathcal{M}_{\text{SM}} + \sum_{i} \frac{c_i}{\Lambda^2} \mathcal{M}_i ,
\end{equation}
\end{linenomath}
where $\mathcal{M}_{\text{SM}}$ is the SM ME, $\mathcal{M}_{i}$ are the MEs associated with the new operators, and $c_i$ are WCs.
A production cross section, either inclusive or differential, is proportional to the square of the total ME and can thus be written as a polynomial of second order in the WCs.
The weight for a given event can then be parameterized with a five-dimensional (5D) quadratic function of the WCs as
\begin{linenomath}
\begin{equation}
\label{eq:param}
 w \Big(\frac{\vec{c}}{\Lambda^2} \Big) = s_{0} + \sum_j s_{1,j} \frac{c_j}{\Lambda^2} + \sum_j s_{2,j} \Big(\frac{c_j}{\Lambda^2}\Big)^2 + \sum_{j} \sum_{k} s_{3,jk} \frac{c_j}{\Lambda^2} \frac{\raisebox{0.3ex}{$c_k$}}{\Lambda^2} ,
\end{equation}
\end{linenomath}
where the sums run over the five WCs, $\vec{c}$ is the set of WC values, and the $s_{0}$, $s_{1}$, $s_{2}$, $s_{3}$ coefficients are associated with: 
the SM amplitude; the interference between SM and EFT amplitudes; EFT amplitudes; and the interference among EFT amplitudes, respectively. 
Although individual coefficients may be negative or null, the sum of all components always yields a physical distribution.

Following the procedure adopted in Ref.~\cite{TOP-19-001}, this analysis makes extensive use of the possibility offered by the $\MGvATNLO$ event generator to assign multiple weights to an event, which represent the infinitesimal contributions from this event to the total cross section at different points in the EFT phase space.
Each simulated event is associated with weights corresponding to different WC values, which are used to build an overdetermined system of equations to solve for all the per-event coefficients---21 in total---of Eq.~(\ref{eq:param}). 
By summing the quadratic functions of all simulated events for a given signal process, one can then evaluate its differential cross section as a function of any quantity, at any point in the 5D EFT phase space.
The parameterized event yields of the signal samples are normalized such that they equal their SM NLO theoretical predictions~\cite{YR4,Sirunyan:2017nbr} when all WCs are set to zero.

\section{Event reconstruction}
\label{sec:reconstruction}
The particle-flow algorithm~\cite{CMS-PRF-14-001} combines information from all subdetectors to reconstruct individual particles in an event, and to identify them either as electrons, muons, photons, charged hadrons, or neutral hadrons.

The candidate vertex with the largest value of summed physics-object $\pt^2$ is taken to be the primary \pp interaction vertex. The physics objects in this context are the jets, clustered using the jet-finding algorithm~\cite{Cacciari:2008gp,Cacciari:2011ma} with the tracks assigned to candidate vertices as inputs, and the negative vector \pt sum of those jets.
Reconstructed lepton candidates are required to have track parameters compatible with the primary vertex as origin.

Electron candidates are reconstructed within the range $\abseta < 2.5$ by combining the energy measurement in the ECAL with the momentum measurement in the tracker~\cite{Khachatryan:2015hwa}. They are required to satisfy $\pt > 7\GeV$, and their identification relies upon an MVA algorithm trained with observables related to the ECAL and tracker measurements.
Electrons originating from photon conversion are efficiently removed by requiring that candidate tracks have at most one missing hit in the innermost tracker layers. 

Muon candidates are reconstructed within the range $\abseta < 2.4$ as tracks in the tracker consistent with either a track or several hits in the muon system, and associated with calorimeter deposits compatible with the muon hypothesis~\cite{Sirunyan:2018}. 
They are required to satisfy $\pt > 5\GeV$, as well as a set of quality criteria designed to reject hadrons punching through the calorimeters and muons produced by in-flight decays of kaons and pions.

Electron and muon candidates satisfying the aforementioned selection criteria are referred to as ``loose leptons''.
Additional selection criteria are applied to select genuine ``prompt'' leptons produced in decays of \PW\ and \PZ\ bosons and leptonic \Pgt decays, while rejecting ``nonprompt'' leptons (NPLs) mainly originating from \cPqb hadron decays, hadron misidentification, and the conversion of photons not produced in the hard scattering interaction. 
Background events containing at least one NPL, arising mostly from $\ttj$ and $\dy$ production, will be referred to as ``NPL background'' throughout this paper. 
The rejection of NPLs is significantly improved by the use of MVA discriminants based on boosted decision trees~\cite{ttH_ML_legacy}.
They take as input several observables related to the lepton and to the jet activity in its vicinity. 
Electron and muon candidates satisfying a selection on the MVA discriminants are referred to as ``tight leptons''.

Jet candidates are reconstructed offline from energy deposits in the calorimeter towers, and clustered using the anti-\kt algorithm~\cite{Cacciari:2008gp, Cacciari:2011ma} with a distance parameter of 0.4. 
They are required to satisfy $\pt > 25\GeV$ and $\abseta < 5$, and must not overlap with any loose lepton within $\Delta R = \sqrt{\smash[b]{(\Delta\eta)^2+(\Delta\phi)^2}} < 0.4$, where $\phi$ is the azimuthal angle.
The contribution from pileup to jet momentum is mitigated by excluding charged hadrons associated with pileup vertices from the clustering. 
The energy of reconstructed jets is corrected for residual pileup effects, and calibrated as a function of jet \pt and $\eta$~\cite{Cacciari:2007fd,Khachatryan:2016kdb}.
We apply the more stringent requirement $\pt > 60\GeV$ to jets reconstructed within the range $2.7 < \abseta < 3$ to suppress calorimeter noise.
Jets passing these selection criteria are categorized into central and forward jets, the former satisfying the condition $\abseta < 2.4$, and the latter $2.4 < \abseta < 5$.
The presence of a high-\pt forward jet in the event, referred to as a recoiling jet, is a characteristic feature of \tZq production that is used in the MVA to isolate the contribution from this process. 
The phase space extension due to the inclusion of forward jets increases the acceptance of the \tZq signal by about 25\% in the trileptonic signal region defined in Section~\ref{sec:selection}.

Jets arising from bottom quarks are identified (\cPqb\ tagged) with the \DeepJet deep neural network algorithm~\cite{BTV-16-002,Bols_2020}, within the ranges $\abseta < 2.4$ in 2016 and $\abseta < 2.5$ in 2017--2018 because of the Phase-1 upgrade of the CMS pixel detector~\cite{pixel_upgrade}. 
Jets passing the medium working point of the algorithm are referred to as ``\cPqb\ jets''. For central jets with $\pt > 30\GeV$, this corresponds to a selection efficiency of about 75\% for jets arising from bottom quarks, and to a misidentification rate for light-quark and gluon jets of about 1\%~\cite{CMS-DP-2018-058}.

The missing transverse momentum vector \ptvecmiss is computed as the negative vector \pt sum of all the particle-flow candidates in an event, and its magnitude is denoted by \ptmiss~\cite{Sirunyan:2019kia}. The \ptvecmiss is modified to account for corrections to the energy scale of the reconstructed jets in the event. 

After the final-state particles have been identified and selected, we combine their information to reconstruct unstable particles that are expected in the topologies of the signal processes (see Fig.~\ref{fig:feynman}). 
These higher-level objects are used for event categorization and to construct powerful observables provided as input to the MVA algorithms presented in Section~\ref{sec:mva}.

The \PZ\ boson candidates are reconstructed from pairs of opposite-sign same-flavor leptons having invariant masses within 15\GeV of the true mass \mz of the \PZ\ boson. 
While this analysis considers events with either three or four leptons, we do not reconstruct tetraleptonic events further since their kinematic information is not exploited in the signal extraction procedure described in Section~\ref{sec:extraction}.
In trileptonic events with multiple \PZ\ boson candidates, only the candidate whose mass is closest to \mz is considered.
The vector \ptvecmiss is associated with the undetected neutrino arising from the leptonic top quark decay. 
The longitudinal neutrino momentum is obtained by applying the \PW\ boson mass constraint and solving an analytical equation. 
The leptonically decaying top quark is then reconstructed using the four-momentum of the neutrino, the remaining selected lepton, and one \cPqb\ jet. 
In case two neutrino solutions or multiple \cPqb\ jets are reconstructed, the combination resulting in a top quark mass closest to $m_{\PQt}$ is chosen. 
The \PW\ boson transverse mass \mtw is reconstructed from the $\PW \to \ell \nu$ decay products and constitutes a powerful observable to discriminate background processes such as \ZZ, $\dy$, and \Zg.
To improve the selection of the \tZq signal, we identify the remaining jet with the largest \pt in the event as the recoiling jet, with a veto on \cPqb\ jets. 
In the rare case that an event contains only \cPqb\ jets, this veto is removed.
The lepton asymmetry $\text{q}_{\ell} \abs{\eta(\ell)}$ is defined as the product of the charge and absolute pseudorapidity of the lepton originating from the top quark decay, and provides additional discrimination power for the \tZq signal.

\section{Event selection and categorization}
\label{sec:selection}
This analysis targets the associated production of top quarks with a \PZ boson, in events where the \PZ boson and at least one top quark decay leptonically. 
The corresponding experimental signatures are characterized by the presence of: three or four prompt leptons; high \ptmiss due to the neutrino(s) from \PW boson decay(s); at least one \cPqb\ jet from top quark decay; and possibly additional light-quark jets, either produced in the decays of top quarks or \PW bosons, or recoiling against the single top quark.

The signal region (SR) targeting trileptonic signal events drives the sensitivity of the analysis, and is denoted by \sr. 
An additional SR targeting tetraleptonic \ttZ events is included, which is denoted by \srttzfour. 
Although this region contains a much smaller amount of data compared with the \sr, it is pure in \ttZ events and thus provides additional sensitivity to operators impacting the cross section of this signal.

\begin{figure} [!hbtp] 
  \centering
  \includegraphics[width=\textwidth]{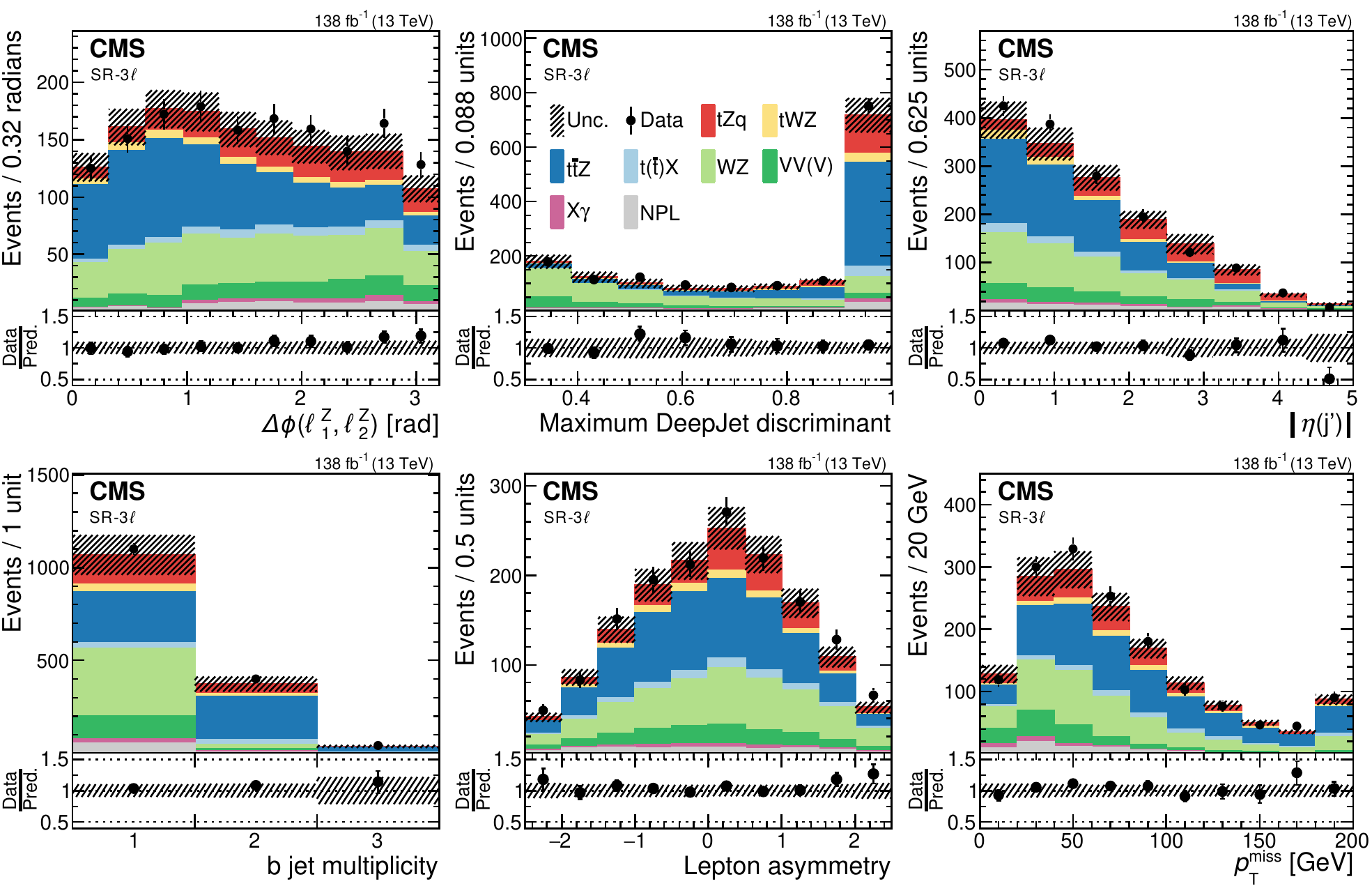}
\caption{Pre-fit data-to-simulation comparisons for several observables in the \sr. 
From left to right and upper to lower, the distributions correspond to: the relative azimuthal angle $\Delta \phi$ between the two leptons from the \PZ boson decay; the maximum \DeepJet discriminant among all selected jets; the absolute pseudorapidity of the recoiling jet; the \cPqb\ jet multiplicity; the lepton asymmetry; and \ptmiss.
The lower panels display the ratios of the observed event yields to their predicted values. 
The NPL background is modeled with the procedure based on control samples in data described in Section~\ref{sec:backgrounds}. 
The hatched band represents the total uncertainty in the prediction. 
Underflows and overflows are included in the first and last bins, respectively.}
\label{fig:controlPlots}
\end{figure}

In the \sr (\srttzfour) we require the presence of three (four) tight leptons with $\pt > 25, 15, 10$ (and 10)\GeV. 
To reject lepton pairs originating from low-mass hadron decays, which are not well modeled by the MC simulation, events containing a pair of loose leptons having an invariant mass below 12\GeV are discarded. 
Since this analysis focuses on the \ttZ interaction, we require the presence of exactly one \PZ\ boson candidate in events entering the SRs.
They are also required to contain at least two jets, and at least one \cPqb\ jet.
Additionally, we require the sum of the lepton charges to be zero in events entering the \srttzfour.
A multiclass classifier is then exploited to divide the \sr into process-specific subregions, as explained in Section~\ref{sec:mva}.
Figure~\ref{fig:controlPlots} shows data-to-simulation comparisons in the \sr for several observables that are most relevant for this classification, before performing a fit to data as described in Section~\ref{sec:extraction} (pre-fit).

The main irreducible background contribution in the \sr arises from the production of two vector bosons, predominantly via the $\WZ \to 3\ell$ process.
The $\ZZ \to 4\ell$ process also contributes, in cases where one lepton fails the lepton identification or is out of detector acceptance.
These events contain on average fewer jets (\cPqb tagged or not) compared with signal processes, 
and are efficiently suppressed by the requirements on the jet and \cPqb\ jet multiplicities.
Background processes with three vector bosons (\WWW, \WWZ, \WZZ, and \ZZZ) are considered as well and represent a minor contribution; together with the \ZZ process, they are denoted by \VVV.
Other processes with top quarks contributing as irreducible backgrounds are \ttH, \ttW, \tHW, \tHq, \ttVV, \ttVH, \ttHH, and \tttt; they are denoted by \tX.
The contribution from processes including a photon produced in the hard scattering interaction is denoted by \Xg, and is dominated by \Zg and \ttg events yielding two prompt leptons plus a photon that undergoes an asymmetric conversion. 
The NPL background is modeled with the procedure based on control samples in data described in Section~\ref{sec:backgrounds}.

Two control regions (CRs) are enriched in the main background processes. They are included in the signal extraction procedure to constrain the uncertainties in the cross sections of these backgrounds from the data and are defined as follows:
\begin{itemize}
 \item A CR enriched in \WZ events (``\WZ CR'') is defined similarly to the \sr, except that events containing one or more \cPqb\ jet(s) are rejected and no requirement on the minimal jet multiplicity is applied. 
The purity in \WZ events is increased by requiring $\ptmiss > 50\GeV$, and the invariant mass of the trileptonic system \mlll to be larger than $\mz + 15\GeV$.
 \item A CR enriched in \ZZ events (``\ZZ CR'') is defined by requiring the presence of exactly four tight leptons that are compatible with two \PZ boson candidates. No requirements on the jet and \cPqb\ jet multiplicities are applied.
\end{itemize}

All regions are orthogonal by construction, and the leptons selected in the CRs must satisfy the same \pt thresholds as used in the SRs.
A summary of the main selection requirements applied in each region is provided in Table~\ref{tab:regions}.

\begin{table}[!hbtp]
  \centering
  \topcaption{
  \label{tab:regions}
  Summary of the main selection requirements applied in each signal or control region. A dash indicates that the selection requirement is not applied.}
  \begin{tabular}{lllll}
    Selection requirement  			    & \sr & \srttzfour        & \WZ CR        & \ZZ CR           \\
    \hline
    Lepton multiplicity                 & $=$3         & $=$4             & $=$3         & $=$4        \\
    $\mlll - \mz$                       & \NA          & \NA              & $>$15\GeV    & \NA          \\
    \PZ boson candidates multiplicity   & $=$1         & $=$1             & $=$1         & $=$2          \\
    Jet multiplicity		            & $\ge$2       & $\ge$2           & \NA          & \NA            \\
    \cPqb\ jet multiplicity	  	        & $\ge$1       & $\ge$1           & $=$0         & \NA             \\
    $\ptmiss$			     	        & \NA          & \NA              & $>$50\GeV    & \NA              \\
  \end{tabular}
\end{table}

\section{Background estimation}
\label{sec:backgrounds}
Irreducible backgrounds containing genuine prompt leptons are reliably estimated from MC simulations and are constrained in dedicated CRs. 
The \Xg background is strongly suppressed by the lepton identification criteria and its remaining contribution is estimated from MC simulations. 
However, the NPL background is known to be much more challenging to model.
Thus, its contribution is estimated using the data-based misidentification probability method~\cite{ttH_ML_legacy}.
This method relies on the selection of samples of events satisfying the same selection criteria as defined in Section~\ref{sec:selection} for the different regions, except that the lepton identification requirements are relaxed. We refer to these samples as the application regions (ARs) of the misidentification probability method. 

Estimates of the NPL background contribution to the different regions are obtained by applying suitably-chosen weights to the corresponding AR events.
These weights quantify the probability for NPLs passing the relaxed identification requirements to be misidentified as tight leptons. 
They are measured in a data sample enriched in events composed uniquely of jets produced via the strong interaction, referred to as multijet events, in which an NPL originating from one of the jets is reconstructed.
These data are collected with single muon (electron) triggers, some of which require the presence of an additional jet with $\pt > 40$ (30)\GeV.
Events are selected if they contain exactly one lepton passing the relaxed identification criteria, plus at least one jet separated by $\Delta R > 0.7$ from the lepton.
The weights are measured separately for electrons and muons, and parameterized with the \pt and $\eta$ of the lepton candidate.
The selected events are divided into ``pass'' and ``fail'' samples, depending on whether the lepton passes or fails the tight identification criteria, respectively.

The contribution from multijet events dominates the fail sample and is estimated directly from the data, after subtracting the small contamination from prompt lepton events (mostly from $\Wj$, $\dy$, diboson, and $\ttj$ production) based on MC simulations.
It is then fit to the data in the pass sample, along with the contribution from prompt lepton events as estimated from the simulations, to determine the number of multijet events in the pass sample.
The misidentification probability in a given category is computed as $w = N_{\text{pass}} / (N_{\text{pass}}+N_{\text{fail}})$, where $N_{\text{pass}}$ and $N_{\text{fail}}$ are the number of multijet events in the pass and fail samples, respectively.
To prevent potential double counting in the NPL background estimation, all tight leptons selected in signal and control regions are matched to their generator-level equivalents in the simulation, and simulated events containing at least one lepton qualifying as nonprompt are discarded.
Further details on the procedure for the estimation of the NPL background can be found in Ref.~\cite{ttH_ML_legacy}.

\section{Multivariate analysis} 
\label{sec:mva}
This analysis makes extensive use of MVA techniques based on neural networks (NNs) to enhance the sensitivity to new phenomena arising from the EFT operators of interest.
Firstly, a multiclass classifier is trained to separate the contributions from the main SM processes in the \sr. It is used to define three subregions enriched in the \tZq, \ttZ, and background processes. 
Secondly, binary classifiers are trained to separate events generated according to the SM from events generated with nonzero WC values. 
The responses of these NNs represent optimal observables that are used for the signal extraction in the \sr subregions.

All trainings are performed with MC simulations in the \sr, using the \Tensorflow~\cite{tensorflow2015-whitepaper} package with the \Keras interface~\cite{keras} and \Adam optimizer~\cite{kingma2017adam}.
The training phase aims to reduce the cross-entropy loss function~\cite{cross-entropy}, and the NN weights are updated using batch gradient descent.
Potential overtraining is mitigated using dropout~\cite{dropout}, L2 regularization~\cite{Tikhonov-regularization}, and early stopping in case the minimized function has reached a stable minimum. 
The number of equidistant bins $N_{\text{bins}}$ and the range of each NN output distribution are adjusted such that the total expected event yield is above one in all bins. 
The bins are numbered from 1 to $N_{\text{bins}}$.

\subsection{Discrimination between SM processes}
\label{sec:mvasm}
The \sr selection criteria are designed to retain a large proportion of signal events while rejecting most background events.
However, basic selection criteria cannot isolate efficiently the rare signals from the overwhelming background processes that yield similar final states.
Moreover, each signal process probes the WCs in unique ways, and the shapes of their kinematic distributions may be impacted differently in the presence of new interactions. To take advantage of this complementarity, it is thus desirable to separate the signal processes from one another.

To this end, we train a multiclass NN classifier, denoted hereafter by ``NN-SM''. 
It is tasked with isolating the \tZq and \ttZ signals from the major \WZ, \tX, and \VVV backgrounds.
The \tWZ signal is not targeted explicitly because it has a comparatively small event yield and is kinematically close to the \ttZ process.
This classifier is trained using the SM signal samples, which represent our best available SM predictions at NLO accuracy for these processes.

The NN-SM includes three hidden layers, each with 100 rectified linear units~\cite{ReLu}, and has three output nodes labeled ``$\tZq$'', ``\ttZ'', and ``Others''. 
The response of the NN-SM is used to divide the \sr into three subregions, each enriched in a particular class of events, which we label accordingly ``$\srtzq$'', ``$\srttz$'', and ``$\srother$''. 
The \textrm{softmax} activation function is used for all output nodes. Their activation values may be interpreted as the probability for a given event to be either \tZq signal, \ttZ signal, or background. 
Since the softmax operation normalizes the output vector to unit magnitude by construction, we assign a given event entering the \sr to either of the orthogonal subregions based on the output node that is most activated by this event.
The set of 33 variables used as input to the \nnsm comprises 12 ``high-level'' variables that are designed to improve the separation between the different process classes. 
The three-momenta (\pt, $\eta$, $\phi$) of the three leading leptons and up to three leading jets in the events, as well as the \cPqb\ tagging discriminants for these jets, are also included and improve the classification performance. All the input variables are listed in Table~\ref{tab:inputs}. 
Their distributions and correlations with other variables are verified to be properly modeled by the simulations. 

\begin{figure} [!hbtp] 
  \centering
  \includegraphics[width=\textwidth]{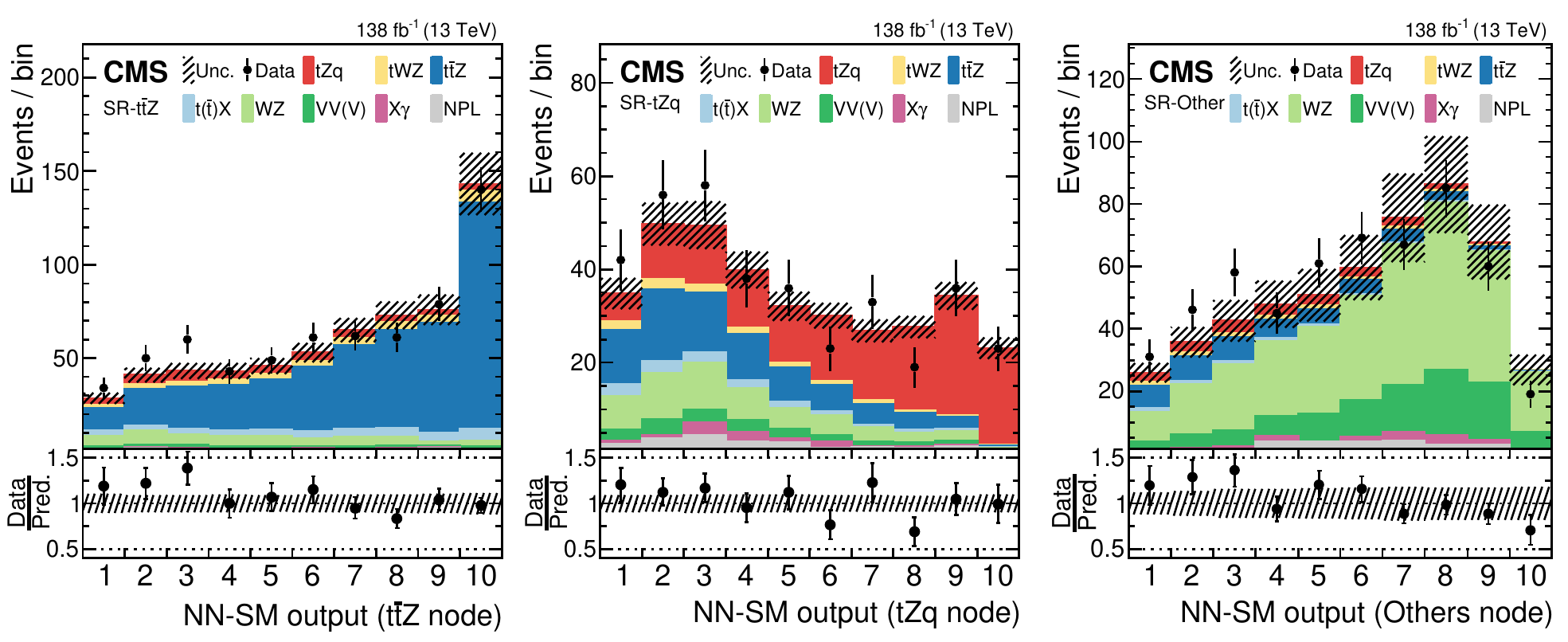}
\caption{Pre-fit data-to-simulation comparisons for the distributions of the \ttZ (left), \tZq (middle), and Others (right) output nodes.
For each distribution, only the events that have their maximum value in the corresponding output node are included.
The lower panels display the ratios of the observed event yields to their predicted values. 
The hatched band represents the total uncertainty in the prediction.}
\label{fig:prefit_nn_sm}
\end{figure}

The pre-fit data-to-simulation comparisons for the distributions of the three output nodes of the \nnsm are shown in Fig.~\ref{fig:prefit_nn_sm}.
For each distribution, only the events that have their maximum value in the corresponding output node are included.

\begin{table}[hbt!]
\centering
\topcaption{
\label{tab:inputs}
Input variables to the \nnsm and to the eight \nnefts.
A dash indicates that the variable is not used.
The three-momentum of an object includes the \pt, $\eta$, and $\phi$ components of its momentum.
The symbol $\ell_{\PQt}$ denotes the lepton produced in the decay of the top quark; $j'$ denotes the recoiling jet; $b$ denotes the \cPqb\ jet associated with the leptonic top quark decay; $(\ell^{\PZ}_{1},\ell^{\PZ}_{2})$ denote the leptons produced in the \PZ boson decay; \costhetaz is the cosine of the angle between the direction of the \PZ boson in the detector reference frame, and the direction of the negatively-charged lepton from the \PZ boson decay in the rest frame of the \PZ boson. 
Other observables are defined in Section~\ref{sec:reconstruction}.}
\cmsTable{
\begin{tabular}{lccccccccc}
Variable & \rot{\nnsm} & \rot{\nnctztzq} & \rot{\nnctzttz} & \rot{\nnctwtzq} & \rot{\nnctwttz} & \rot{\nncpqttzq} & \rot{\nncpqtttz} & \rot{\nnalltzq}  & \rot{\nnallttz} \\
\hline\\[-2.2ex]
\zpt                            & \NA        & \checkmark & \checkmark & \checkmark & \checkmark & \checkmark & \checkmark & \checkmark & \checkmark \\
$\eta(Z)$						& \checkmark & \checkmark & \checkmark & \NA        & \NA        & \checkmark & \NA        & \NA        & \checkmark \\
$\Delta\phi(\ell^{\PZ}_{1}\ell^{\PZ}_{2})$ 
                                & \checkmark & \checkmark & \checkmark & \checkmark & \checkmark & \checkmark & \checkmark & \checkmark & \checkmark \\
$\pt(\PQt)$                    & \checkmark & \checkmark & \checkmark & \NA        & \checkmark & \checkmark & \NA        & \checkmark & \checkmark \\
$\eta(\PQt)$                   & \NA        & \checkmark & \checkmark & \checkmark & \checkmark & \checkmark & \NA        & \NA        & \checkmark \\
$m(\PQt,\PZ)$       			& \NA        & \NA        & \NA        & \NA        & \NA        & \NA        & \NA        & \NA        & \NA         \\
$\abs{\eta(j')}$				& \checkmark & \NA        & \NA        & \NA        & \NA        & \NA        & \NA        & \checkmark & \NA         \\
$\pt(j')$                       & \checkmark & \checkmark & \NA        & \checkmark & \NA        & \NA        & \NA        & \NA        & \NA         \\
$\Delta R(b,\ell_{\PQt})$      & \NA        & \checkmark & \NA        & \checkmark & \NA        & \NA        & \NA        & \NA        & \NA         \\
$\Delta R(j',\ell_{\PQt})$     & \checkmark & \NA        & \NA        & \NA        & \NA        & \NA        & \NA        & \NA        & \NA         \\
$\Delta R(\PQt,\PZ)$       	& \NA        & \checkmark & \checkmark & \checkmark & \NA        & \checkmark & \NA        & \NA        & \checkmark \\
$\Delta \eta(\PZ,j')$       	& \NA        & \checkmark & \NA        & \NA        & \NA        & \NA        & \NA        & \checkmark & \NA         \\
$\Delta R$ between \PQt and the closest lepton       
                                & \NA        & \checkmark & \NA        & \checkmark & \NA        & \NA        & \NA        & \NA        & \NA         \\
$\Delta R$ between $j'$ and the closest lepton      
                                & \NA        & \NA        & \NA        & \NA        & \NA        & \NA        & \NA        & \checkmark & \NA         \\
$\mlll$						    & \checkmark & \NA        & \NA        & \NA        & \checkmark & \NA        & \checkmark & \NA        & \checkmark \\
$\mtw$                          & \checkmark & \checkmark & \checkmark & \NA        & \NA        & \NA        & \NA        & \NA        & \checkmark \\
$\ptmiss$         	            & \checkmark & \NA        & \NA        & \NA        & \NA        & \NA        & \NA        & \NA        & \NA         \\
Lepton asymmetry         		& \checkmark & \NA        & \NA        & \checkmark & \checkmark & \NA        & \NA        & \checkmark & \NA  \\
\costhetaz					    & \NA        & \NA        & \checkmark & \NA        & \NA        & \checkmark & \NA        & \NA        & \checkmark \\
Max. \pt among jet pairs        & \NA        & \NA        & \NA        & \NA        & \NA        & \NA        & \checkmark & \NA        & \checkmark \\
Max. \DeepJet discriminant      & \checkmark & \NA        & \NA        & \NA        & \NA        & \NA        & \NA        & \NA        & \NA         \\
\cPqb\ jet multiplicity         & \checkmark & \NA        & \NA        & \NA        & \NA        & \NA        & \NA        & \NA        & \NA         \\ [\cmsTabSkip]
Three-momenta of the three leading leptons                   
                                & \checkmark & \NA        & \NA        & \NA        & \NA        & \NA        & \NA        & \NA        & \NA         \\
Three-momenta of the three leading jets 
                                & \checkmark & \NA        & \NA        & \NA        & \NA        & \NA        & \NA        & \NA        & \NA         \\
\DeepJet discriminants of the three leading jets 
                                & \checkmark & \NA        & \NA        & \NA        & \NA        & \NA        & \NA        & \NA        & \NA         \\ [\cmsTabSkip]
Number of variables             & 33         & 11         & 8          & 8          & 6          & 7          & 4          & 7          & 10          \\
\end{tabular}
}
\end{table}

\subsection{Discrimination between the SM and EFT scenarios}
\label{sec:mvaeft}
In a second step, we leverage the EFT parameterization of the signal event weights to train NNs tasked with separating events generated either according to the SM, or according to EFT scenarios in which at least one WC is nonzero. These classifiers are denoted by ``\nnefts'' hereafter.
They are used to design observables with optimal sensitivity to new effects arising from the targeted operators.
These techniques are based on ideas developed in the context of likelihood-free inference techniques, 
which are described extensively in Ref.~\cite{eft_ml1}.

Classification algorithms were already used in previous analyses adopting an EFT framework. 
For instance, Ref.~\cite{DHondt:2018cww} takes advantage of a multiclass classifier to separate SM events from events corresponding to the pure EFT contribution to the targeted signal process, as well as to distinguish among different classes of operators. 
However, the interference between EFT operators and the SM amplitude---and among EFT amplitudes---were always either absent, or were voluntarily neglected in the design of these algorithms.
This is because the shapes of the kinematic distributions due to pure EFT contributions are independent of the WCs, 
which makes it possible to train an algorithm on simulated samples whose kinematic properties are unambiguously defined.
On the contrary, interference terms introduce a dependence of the shapes of the kinematic distributions on the WCs, 
which cannot be dealt with efficiently using the most commonly used, problem-specific algorithms. 
This is the first time that an LHC analysis uses machine-learning techniques that account for interference in the training phase to target EFT effects. 

The \nnefts are binary NN classifiers trained to discriminate between the SM and EFT scenarios.
We segment this challenging task by targeting individual WCs with separate trainings, and by targeting the \tZq and \ttZ signals separately, since both processes lead to significantly different event topologies.
As anticipated, we find that for WC values close to the current exclusion limits, the \opqm and \opt operators have negligible impacts on the shapes of the kinematic distributions of the \tZq and \ttZ processes~\cite{Brivio:2019ius}. 
We therefore do not train dedicated classifiers for these cases, but we rely on the sensitivity of this analysis to the signal cross sections to constrain these WCs.
Additional classifiers, referred to explicitly as ``\nnall'', are trained with events sampled over the phase space spanned by \ctz, \ctw, and \cpqt, and will be used to constrain multiple WCs simultaneously. 
Consequently, eight \nnefts are trained and labeled according to the WC(s) and signal process that they target: \nnctztzq, \nnctzttz, \nnctwtzq, \nnctwttz, \nncpqttzq, \nncpqtttz, \nnalltzq, and \nnallttz.
The output distributions of the NNs targeting either the \tZq or \ttZ process are ultimately used to extract the signal in the \srtzq or \srttz subregions, respectively, as described in Section~\ref{sec:extraction}.
Separate sets of statistically independent samples are used to perform the trainings and the signal extraction.

Each \nneft classifier is trained on a range of distinct EFT scenarios. 
The corresponding subsets of training events are sampled uniformly over the range $[-5,5]$ for \ctz and \ctw, and $[-10,10]$ for \cpqt. 
These ranges for the training hypotheses are chosen to be broader than the existing direct constraints on the corresponding WCs. This avoids biasing the trainings on preexisting results and may help the classifiers learn the characteristic features associated with new interactions, since these features are more prominent at larger WC values. The exact definition for these ranges has only a minor impact on the classification performance.
We rely upon the capability of these NNs to learn from training events corresponding to various hypotheses, and to interpolate between these events to construct an abstract model optimized to separate reconstructed events that are either SM- or EFT-like.

The \nnefts include three hidden layers, each with 50 rectified linear units.
The sigmoid activation function is used for the single output node. 
The sets of input variables are defined independently for each classifier and are listed in Table~\ref{tab:inputs}. 
The choice of input variables is informed by generator-level studies of their expected sensitivities, as well as by their correlations, importance rankings, and modeling.
Multiple tests were performed using different architectures, activation functions, optimizers, and learning and dropout rates, and the configuration performing best on a set of simulated events not used during the training phase was retained.

\section{Systematic uncertainties}
\label{sec:systematics}
Many different sources of systematic uncertainty may alter the event yields, the shapes of the observables used in the final fits to data, or both. 
They are treated as correlated between the different regions and data-taking periods, unless stated otherwise.
A summary of the sources of systematic uncertainty included in the measurements is provided in Table~\ref{tab:syst}.

\subsection{Experimental uncertainties}

\begin{itemize}
 \item \textit{Integrated luminosity} The integrated luminosities of the 2016, 2017, and 2018 data-taking periods are individually known with uncertainties in the 1.2--2.5\% range~\cite{CMS-LUM-17-003,LUM-17-004,LUM-18-002}. The total Run~2 integrated luminosity has an uncertainty of 1.6\%. 
 \item \textit{Trigger efficiency} The combination of single-, double-, and triple-lepton trigger algorithms used in this analysis achieves a trigger efficiency of nearly 100\% for events passing the SR or CR selection criteria, both in data and simulation, and therefore no correction is applied. 
 However, a 2\% uncertainty is assigned to the event yields of the MC samples independently for each year to account for the statistical uncertainty in the efficiency measurement in data. 
 \item \textit{Pileup} Event weights are corrected to account for differences in the pileup distributions between data and MC samples, before any event selection is applied. To estimate the corresponding uncertainty, the \pp inelastic cross section is varied by $\pm 4.6\%$~\cite{pileup} and the variations are propagated to the event weights.
 \item \textit{Lepton identification and isolation efficiencies} Event weights are corrected to account for differences in the lepton identification and isolation efficiencies between data and simulation, as a function of the lepton $\pt$ and $\eta$. These corrections are varied independently for electrons and muons, within the uncertainties in the corresponding measurements.
 \item \textit{\cPqb\ jet identification efficiency} Event weights are corrected to make the simulated distribution of the \DeepJet discriminant match that measured in the data. These corrections are parameterized with the jet \pt and $\eta$, and the corresponding uncertainties are measured in $\ttbar$+jets and multijet events~\cite{BTV-16-002}. Several sources of uncertainty are considered, which are related to the contaminations and statistical fluctuations in the measurement regions of the corrections. Uncertainties that are statistical in nature are uncorrelated between the years.
 \item \textit{Jet energy and missing transverse momentum} The jet energy scale is measured with an uncertainty of a few percent, depending on the jet \pt and $\eta$, in $\dy \to \Pe\Pe$, $\dy \to \Pgm\Pgm$, $\Pgg$+jets, dijet, and multijet events~\cite{Khachatryan:2016kdb}. The resulting effect on signal and background distributions is evaluated by varying the energies of jets in simulated events within their uncertainties, recalculating all kinematic observables, and reapplying the event selection criteria. This effect is split between several subsources of uncertainty that are either correlated or uncorrelated between the years. The uncertainties in the jet energy resolution and in the vector \ptvecmiss are evaluated in a similar way, and have smaller effects than the uncertainty in the jet energy scale. These uncertainties are treated as uncorrelated between the years.
 \item \textit{L1 ECAL trigger inefficiency} A gradual shift in the timing of the inputs of the ECAL L1 trigger in the forward region ($\abseta>2.4$) constitutes an additional source of inefficiency in the 2016 and 2017 data-taking periods~\cite{Sirunyan:2020zal}. Dedicated event weights are applied to simulated events to account for this effect and are varied within their uncertainties.
\end{itemize}

\subsection{Theoretical uncertainties}
Several sources of theoretical uncertainty related to the modeling of the signal samples are included.
Except for the uncertainties in the SM cross sections, these uncertainties impact both the event yields through changes in signal acceptance, and the shapes of the observables used for the signal extraction.

\begin{itemize}
 \item \textit{PDF and QCD coupling} The uncertainty in the PDF prediction is estimated by reweighting signal events according to 100 eigenvectors of the \NNPDFThreeOne PDF set. The total uncertainty is measured following the PDF4LHC recommendations~\cite{Butterworth:2015oua}. The value of the QCD coupling \alpS is also varied independently within its uncertainty. 
  \item \textit{Renormalization and factorization scales} The effect of missing higher-order corrections on the distributions of the discriminating observables is estimated by varying the renormalization and factorization scales ($\mu_{\mathrm{R}}$ and $\mu_{\mathrm{F}}$) up and down by a factor of two with respect to their nominal values, avoiding cases in which the two variations are done in opposite directions.
 \item \textit{SM cross sections} We assign uncertainties to the SM cross sections of the signal processes based on theoretical predictions computed at NLO accuracy. 
We sum in quadrature the uncertainties in the PDF, the renormalization scale, and the factorization scale. This results in a normalization-only uncertainty of ${}^{{+}10.0}_{{-}11.6}\%$ for the \ttZ process~\cite{YR4}, and of $\pm 3.3\%$ for the \tZq process~\cite{Sirunyan:2017nbr}. As there is yet no experimental evidence for the existence of the \tWZ process, we assign a more conservative uncertainty of 20\% to its cross section.
 \item \textit{Parton shower} The uncertainty in the PS simulation is estimated by varying the renormalization scale up and down by a factor of two with respect to its nominal value, independently for initial- and final-state radiation (ISR and FSR). The effect of the ISR uncertainty is uncorrelated between the signal processes.
 \item \textit{Additional radiation} Comparisons between the SM \tZq sample (generated at NLO accuracy) and the \tZq sample used for the signal extraction (generated at LO without an extra final-state parton) show good agreement, except for a discrepancy in the jet multiplicity distribution that is not covered by other uncertainties.
We include this effect by adding a systematic uncertainty based on these differences in bins of jet multiplicity, which only affects the shapes of observables for the \tZq signal.
\end{itemize}

\subsection{Background uncertainties}

\begin{itemize}
 \item \textit{Irreducible backgrounds} Uncertainties are assigned to the cross section predictions of all irreducible background processes.
We assign uncertainties of 10\% to the cross sections of the \WZ~\cite{wz_cms}, \VVV~\cite{zz_cms,vvv_cms}, and \Xg~\cite{xgamma_cms,ttgamma_atlas} processes, and of 20\% to those of \tX processes~\cite{ttH_ML_legacy,ttw_cms}. An additional extrapolation uncertainty of 6\% is applied to \WZ events that enter any of the signal regions, to account for differences in the multiplicity of heavy-flavor jets between the \WZ CR and SRs, as estimated from simulations.
 \item \textit{Misidentified-lepton probability estimation} The misidentification probability measurements described in Section~\ref{sec:backgrounds} are affected by several sources of uncertainty related to the limited amount of data in the measurement region, the subtraction of the prompt lepton contamination, and potential differences in kinematic properties of NPL candidates between the measurement and application regions. The weights obtained with this procedure are varied independently for electrons and muons within each uncertainty. Since these uncertainties were estimated in the context of Ref.~\cite{ttH_ML_legacy}, we apply an additional 30\% uncertainty to the event yield of the NPL background to account for differences in the definitions of the application regions of this analysis.
\end{itemize}

{\renewcommand{\arraystretch}{1.05}
\begin{table}[!th]
\caption{Summary of the different sources of systematic uncertainty included in the measurements. 
The first column indicates the source of the uncertainty. The second column indicates whether the source affects the event yields, the shapes of the observables, or both. In the third column, the symbols~``\checkmark'' and~``\NA'' indicate 100\% and 0\% correlations between the data-taking periods, respectively.
}
\label{tab:syst}
\centering
\begin{tabular}{cllc}
 & Source & Type  & Correlation \\ \hline
 
\multirow{2}{*}[-7ex]{\rotatebox[origin=c]{90}{Experimental}} & Integrated luminosity & Yield & Partial \\ & Trigger efficiency & Yield & \NA \\ & Pileup & Both & \checkmark \\ & Lepton identification and isolation & Both & \checkmark \\ & \cPqb\ tagging & Both & Partial \\ & Jet energy scale & Both & Partial \\ & Jet energy resolution & Both & \NA \\ & Missing transverse momentum & Both & \NA \\ & L1 ECAL inefficiency & Both & \checkmark \\ [\cmsTabSkip] 

\multirow{2}{*}[-4ex]{\rotatebox[origin=c]{90}{Theoretical}} & PDF & Both & \checkmark \\ & \alpS & Both & \checkmark \\ & ME scales $\mu_{\mathrm{R}},\mu_{\mathrm{F}}$ & Both & \checkmark \\ & Signal SM cross sections & Yield & \checkmark \\ & ISR and FSR & Both & \checkmark \\ & Additional radiation & Shape & \checkmark \\ [\cmsTabSkip]

\multirow{2}{*}[-3ex]{\rotatebox[origin=c]{90}{Backgrounds}} & WZ normalization & Yield & \checkmark \\ & \VVV normalization & Yield & \checkmark \\ & \tX normalization & Yield & \checkmark \\ & \Xg normalization & Yield & \checkmark \\ & NPL normalization & Yield & \checkmark \\ & NPL misidentification probabilities & Both & \checkmark \\       
\end{tabular}
\end{table}
}

\section{Signal extraction} 
\label{sec:extraction}
The statistical analysis of the data relies upon the construction of a binned likelihood function $\mathcal{L}(\text{data} | \theta, \nu)$ that incorporates all the necessary information regarding the different model parameters. 
The WCs constitute the parameters of interest, denoted by $\theta$,  
and all the systematic uncertainties defined in Section~\ref{sec:systematics} are treated as nuisance parameters, denoted by $\nu$.
The parameterization of the signal event weights with the WCs makes it possible to vary the event yields in all bins, to describe any point in the parameter space spanned by the WCs.
Statistical fluctuations due to the finite number of simulated events are incorporated into the likelihood function via the approach described in Refs.~\cite{MCStats,nuisance1}. 
The compatibility of given values of the model parameters with the data is quantified by performing a maximum likelihood fit in which the negative log-likelihood (NLL) function $-2 \ln(\mathcal{L})$ is minimized and the nuisance parameters are profiled, simultaneously in the six regions (four SRs and two CRs) for the three years. 

We perform a one-dimensional (1D) likelihood scan for each WC by repeating the maximum likelihood fit in steps of that WC, while the four other WCs are fixed to their SM values of zero. 
Any point in the scan is described by the difference $-2 \Delta \ln(\mathcal{L})$ between its NLL value and that of the global minimum. 
The boundaries of the 68 and 95\% confidence level (\CL) confidence intervals are defined as the intersections of the NLL function with values of 1 and 3.84, respectively.
Five individual 1D likelihood scans are thus required to construct confidence intervals for all five WCs.
In addition, we perform two-dimensional (2D) likelihood scans in which two WCs are scanned simultaneously, while the three remaining WCs are fixed to zero; the boundaries of the 68 and 95\% CL confidence intervals are defined as the intersections of the NLL function with values of 2.30 and 5.99, respectively. 
The corresponding plots shown in Section~\ref{sec:results} illustrate the correlations between pairs of WCs.
Finally, we also perform a single 5D maximum likelihood fit in which all five WCs are treated as parameters of interest, and we compute the corresponding 95\% \CL confidence intervals. 

Different observables are used to extract the signal in the \srtzq and \srttz, depending on the number of parameters of interest (one, two, or five) and on the WCs that are probed, while the observables used in all other regions are the same for all fits.
In the \srother we use the \mtw distribution since it provides good separation between the main background processes in this region. 
We perform simple counting experiments in the \srttzfour, \WZ CR, and \ZZ CR regions, as they are already very pure in their target processes and do not require any further discrimination.
The distributions of the different NN classifiers are used to extract the signal in the \srtzq and \srttz subregions, which drive almost entirely the sensitivity of this analysis.
For the 1D fits to the \ctz, \ctw, and \cpqt coefficients, we use the distributions of the dedicated \nneft classifiers. 
For the 1D fits to the \cpqm and \cpt coefficients, we use the distributions of the \tZq and \ttZ output nodes of the \nnsm classifier.
The good discrimination power of the latter observables between the relevant SM processes improves the sensitivity to the signal event yields.
The 2D and 5D fits are performed to the distributions of the \nnall classifiers, which are trained to learn the features associated with several EFT operators simultaneously.
The observables used in the different fits are listed for each region in Table~\ref{tab:observables}.

{\renewcommand{\arraystretch}{1.3}
\begin{table}[!h]
\caption{Observables used in each region for the different fits. The \nnsm is trained to separate different SM processes, while the other NNs are trained to identify new effects arising from one or more EFT operators, as described in Section~\ref{sec:mva}.}
\label{tab:observables}
\centering
\cmsTable{
\begin{tabular}{l{c}@{\hspace*{5pt}}llcccc}
  \multicolumn{1}{l}{Fit configuration}&& \multicolumn{6}{c}{Region} \\ \cline{3-8}
 && $\srtzq$   & $\srttz$  & \srother & \srttzfour   & CR $\WZ$ & CR $\ZZ$\\[\cmsTabSkip]
1D $\ctz$              && \nnctztzq  & \nnctzttz & \multirow{6}{*}{$\mtw$} & \multicolumn{3}{c}{\multirow{6}{*}{\centering Counting experiments}} \\
1D $\ctw$              && \nnctwtzq  & \nnctwttz &                         &                             &                         &                             \\ 
1D $\cpqt$             && \nncpqttzq & \nncpqtttz     &                         &                             &                         &                             \\
1D $\cpqm$             && \nnsm (\tZq node)      & \nnsm (\ttZ node)     &                         &                             &                         &                             \\ 
1D $\cpt$              && \nnsm (\tZq node)     & \nnsm (\ttZ node)     &                         &                             &                         &                             \\ 
2D and 5D              && \nnalltzq  & \nnallttz &                         &                             &                         &                             \\
\end{tabular}
}
\end{table}
}

\section{Results}
\label{sec:results}
\begin{figure} [!hbtp] 
  \centering
  \includegraphics[width=\textwidth]{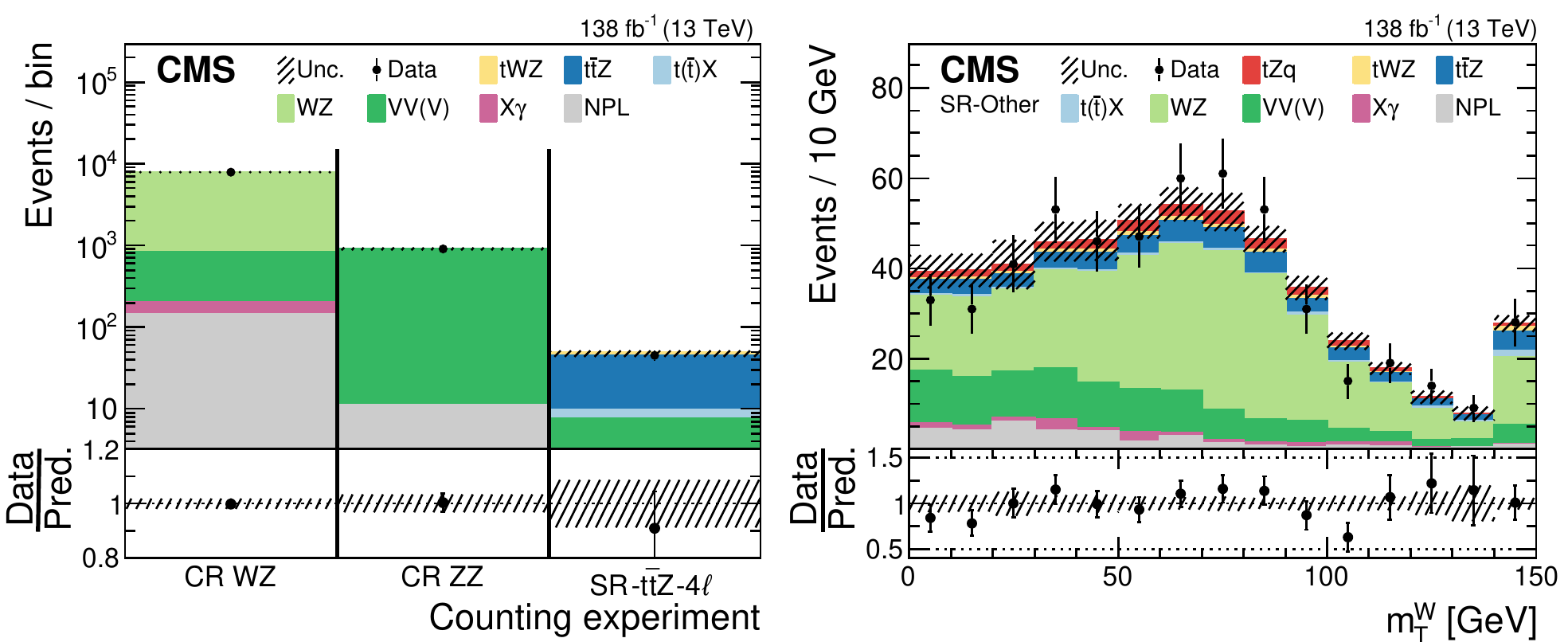}
\caption{Post-fit data-to-simulation comparisons for the distributions that are common to all fits, corresponding to counting experiments in the CRs and \srttzfour (left), and to the \mtw observable in the \srother (right), after the 5D fit. The lower panels display the ratios of the observed event yields to their post-fit expected values. Overflows are included in the last bin of the right figure.}
\label{fig:postfit_common}
\end{figure}

\begin{figure} [!hbtp] 
  \centering
  \includegraphics[width=\textwidth]{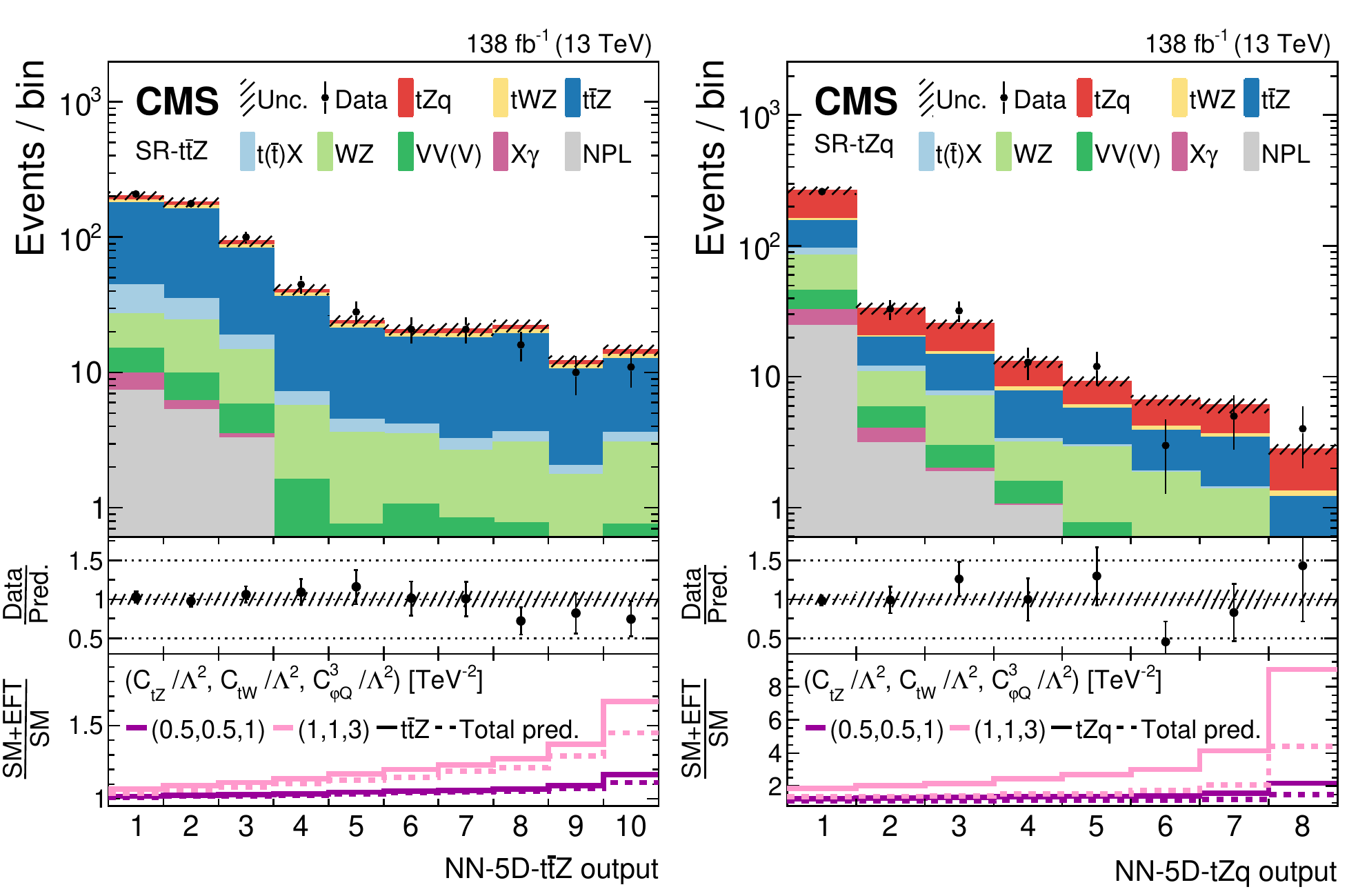}
  \includegraphics[width=\textwidth]{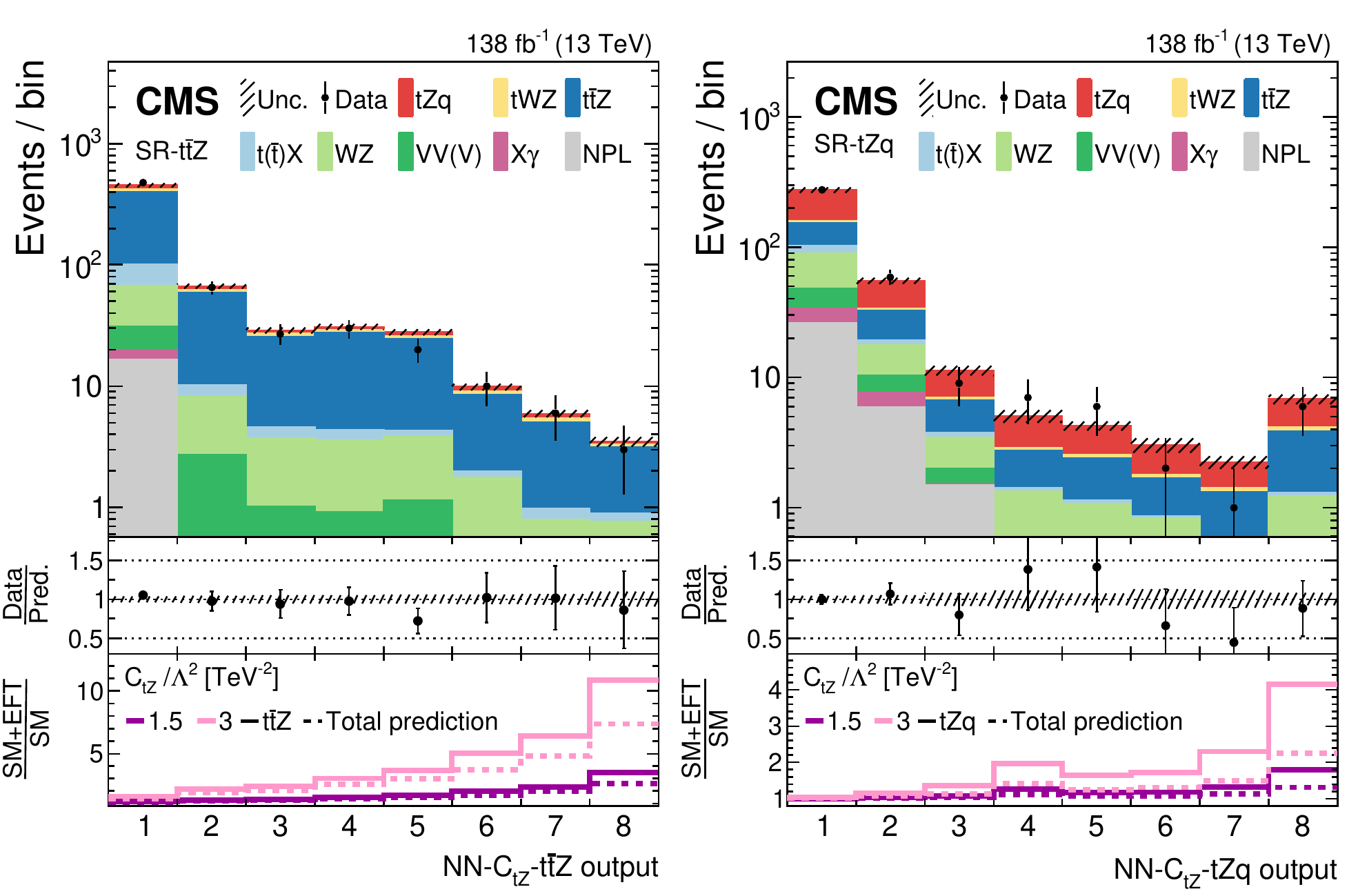}
\caption{Post-fit data-to-simulation comparisons for the distributions used in the \srttz (left) and \srtzq (right), for the 5D fit (upper) and for the 1D fit to \ctz (lower). 
The middle panels display the ratios of the observed event yields to their post-fit expected values.
For each region, the lower panel shows the change of the event yield in each bin with respect to the SM post-fit expectation for two benchmark EFT scenarios, 
both for the main signal process in the region (thick lines) and for the total prediction (dashed lines).}
\label{fig:postfit_1}
\end{figure}

\begin{figure} [!hbtp] 
  \centering
  \includegraphics[width=\textwidth]{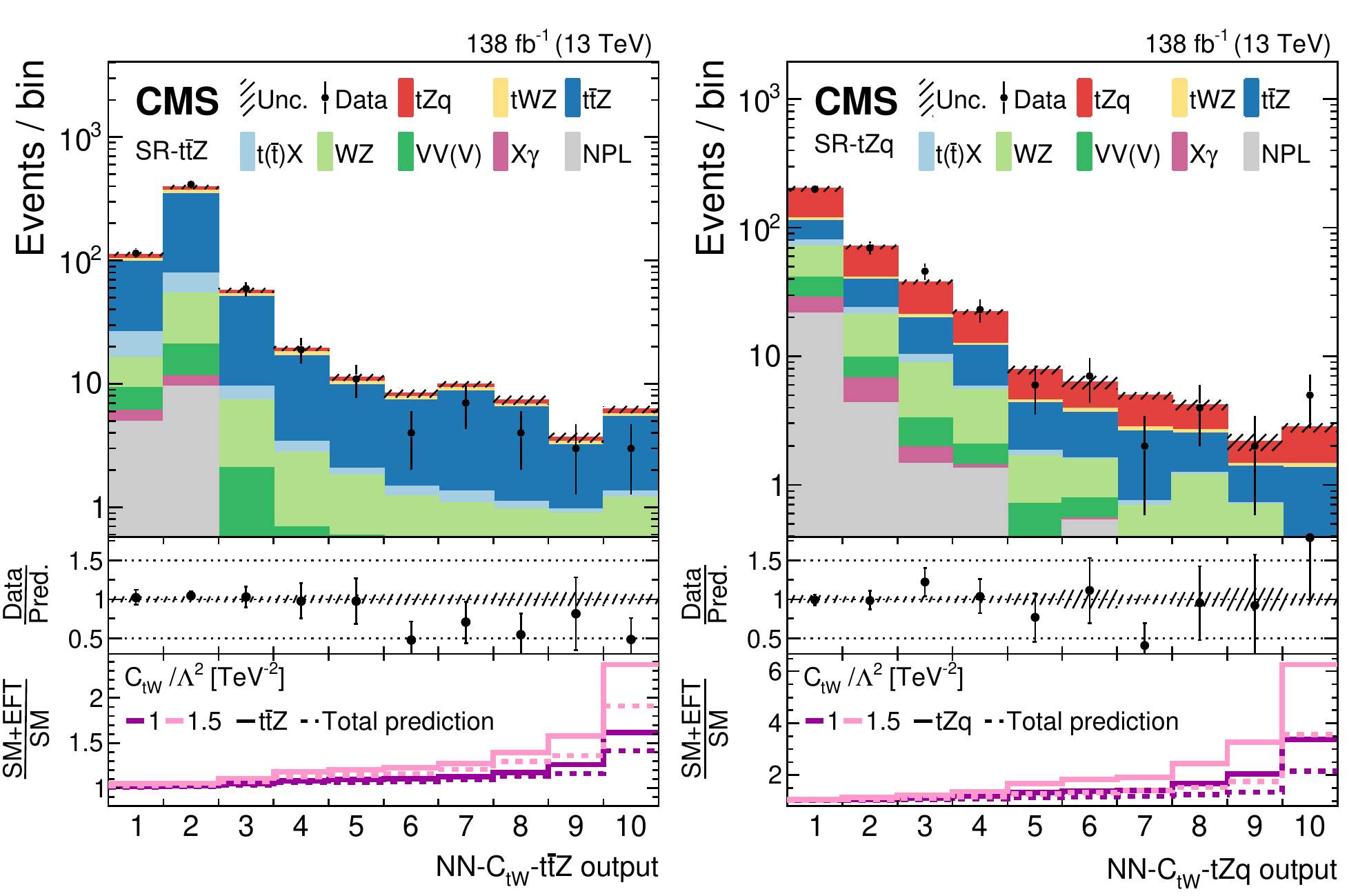}
  \includegraphics[width=\textwidth]{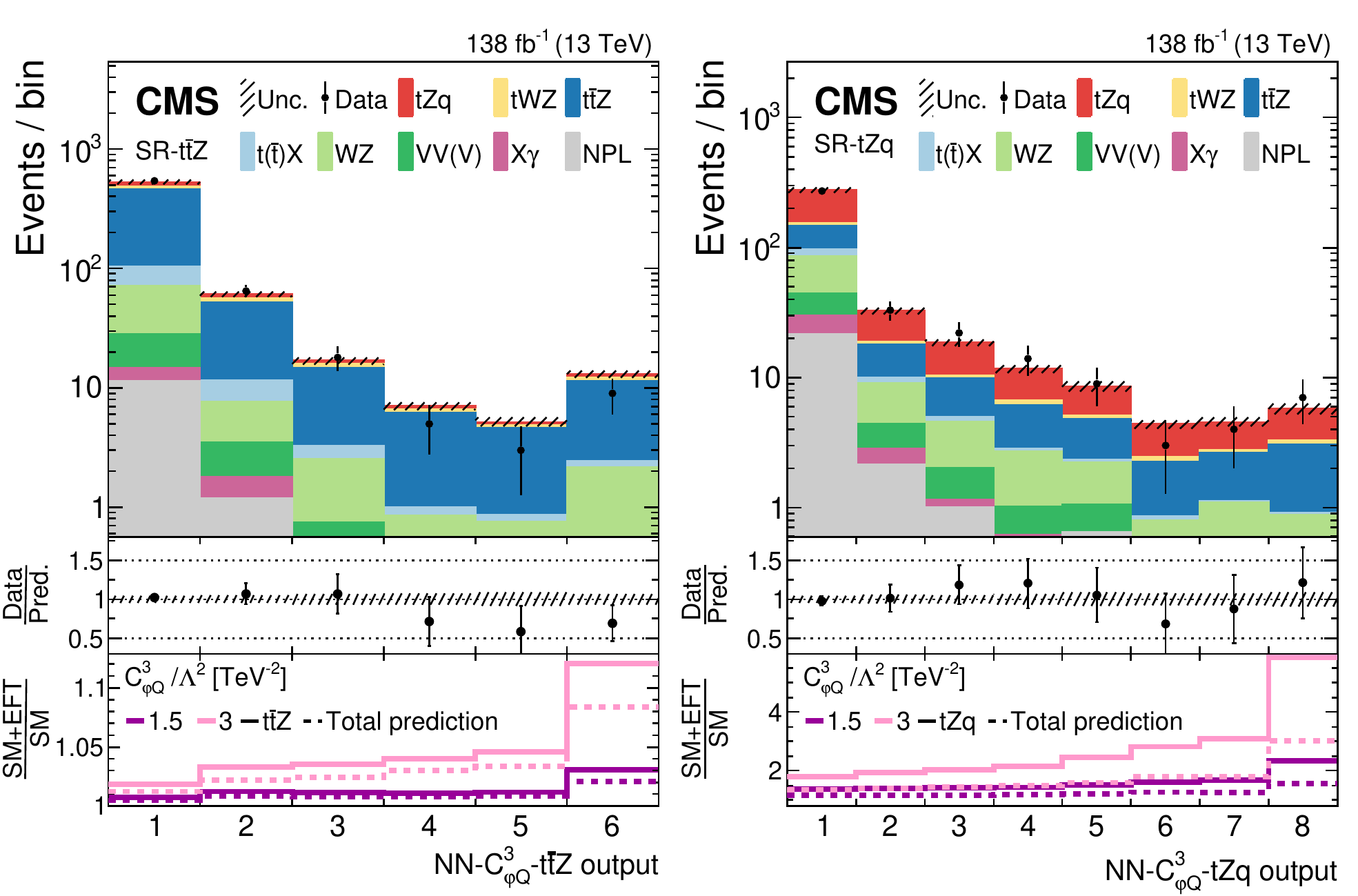}
\caption{Post-fit data-to-simulation comparisons for the distributions used in the \srttz (left) and \srtzq (right), for the 1D fits to \ctw (upper) and to \cpqt (lower). 
The middle panels display the ratios of the observed event yields to their post-fit expected values. 
For each region, the lower panel shows the change of the event yield in each bin with respect to the SM prediction for two benchmark EFT scenarios, 
both for the main signal process in the region (thick lines) and for the total prediction (dashed lines).}
\label{fig:postfit_2}
\end{figure}

The distributions that are common to all fits are shown in Fig.~\ref{fig:postfit_common}, after performing the fit (post-fit) in which all five WCs are treated as free parameters. 
The post-fit distributions of the observables used in the \srtzq and \srttz for the 5D fit and for the 1D fits to the \ctz, \ctw, and \cpqt coefficients, are shown in Figs.~\ref{fig:postfit_1}--\ref{fig:postfit_2}.
The post-fit distributions of the response of the \nnsm in the \srtzq and \srttz, which are used for the 1D fits to the \cpqm and \cpt coefficients, are not shown because they do not exhibit any shape discrimination for the considered EFT scenarios (as explained in Section~\ref{sec:mvaeft}), and because they are very similar to the corresponding pre-fit  distributions shown in Fig.~\ref{fig:prefit_nn_sm}.
We observe good agreement between data and simulation in all distributions, and illustrate the separation powers of the different \nneft classifiers for benchmark EFT scenarios.

The expected and observed 95\% \CL confidence intervals obtained from the 1D and 5D fits are provided in Table~\ref{tab:intervals}.
All intervals include the SM expectations of zero for the WC values.
The optimal combination of the available kinematic information by the NN-EFT classifiers significantly reduces the widths of the confidence intervals, as estimated by repeating the fits when performing simple counting experiments in all regions.
The widths of the expected 1D confidence intervals for \ctw, \ctz, and \cpqt are reduced by about 55, 40, and 20\%, respectively. 
Those for \cpqm and \cpt change only marginally, as anticipated since these WCs mainly affect the signal cross sections. 
The widths of the expected 5D confidence intervals for \ctw, \cpqt, \ctz, \cpqm, and \cpt are reduced by about 70, 70, 60, 35, and 20\%, respectively. 
The corresponding observed intervals exhibit similar or slightly larger improvements. 
This indicates that leveraging kinematic information becomes all the more important when aiming to constrain multiple WCs simultaneously, as it helps with disentangling effects from different EFT operators that have an interplay. 

The impacts from different groups of sources of systematic uncertainty on each individual WC are listed in Table~\ref{tab:impact_wcs}.
We find that the uncertainties in the \ctz and \ctw coefficients are primarily due to the limited number of data events in the most sensitive bins of the \sr distributions.
Conversely, the measurements of the \cpqt, \cpqm, and \cpt coefficients are limited by systematic uncertainties, with those in the signal and background cross sections being dominant. 
As anticipated, since we mostly rely on our sensitivity to the \ttZ event yield to constrain the \cpqm and \cpt coefficients, the corresponding confidence intervals are significantly impacted by the uncertainty assigned to the SM cross section of the \ttZ process.

\begin{figure}[hp!]
  \centering
  \includegraphics[width=0.49\textwidth]{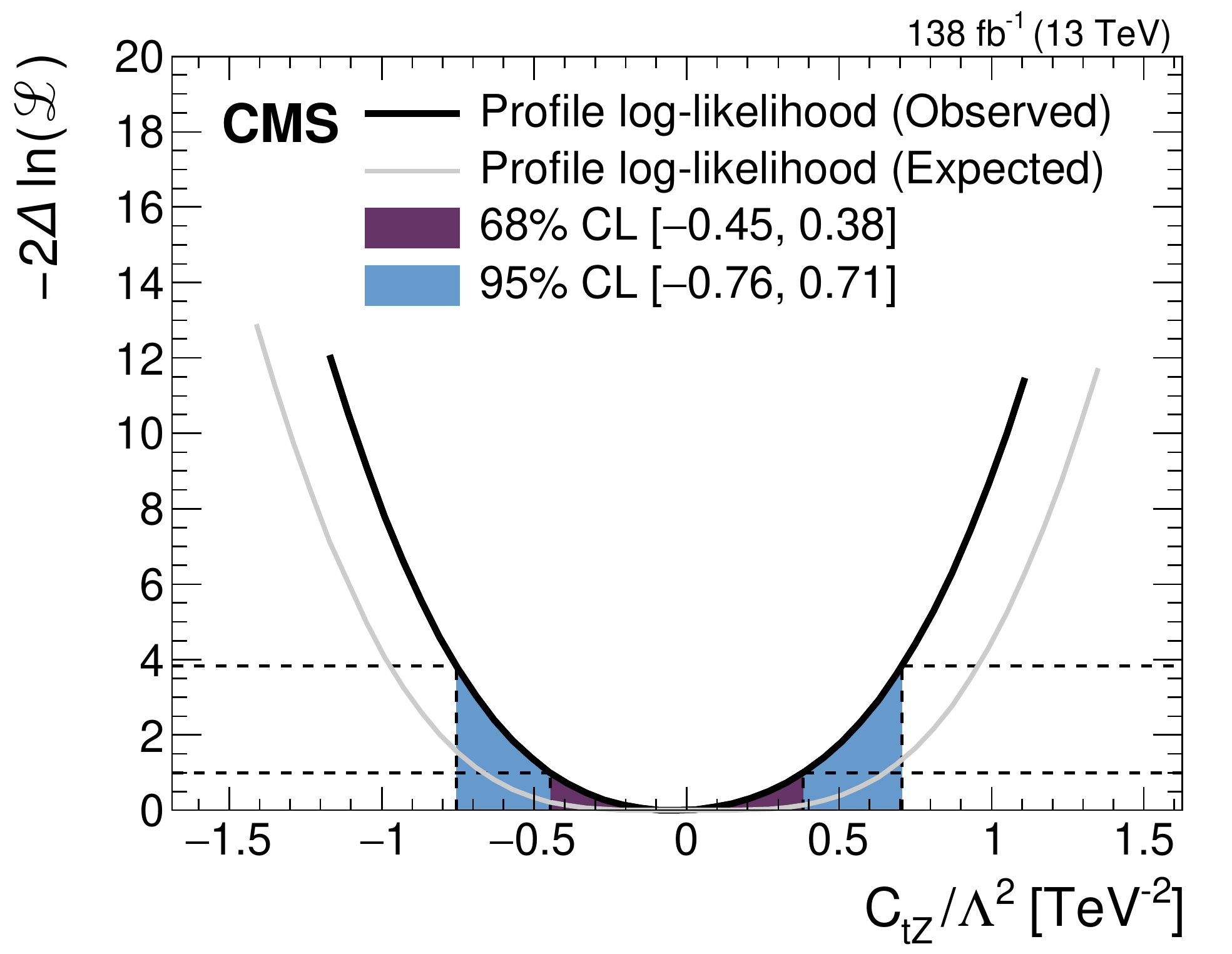} 
  \includegraphics[width=0.49\textwidth]{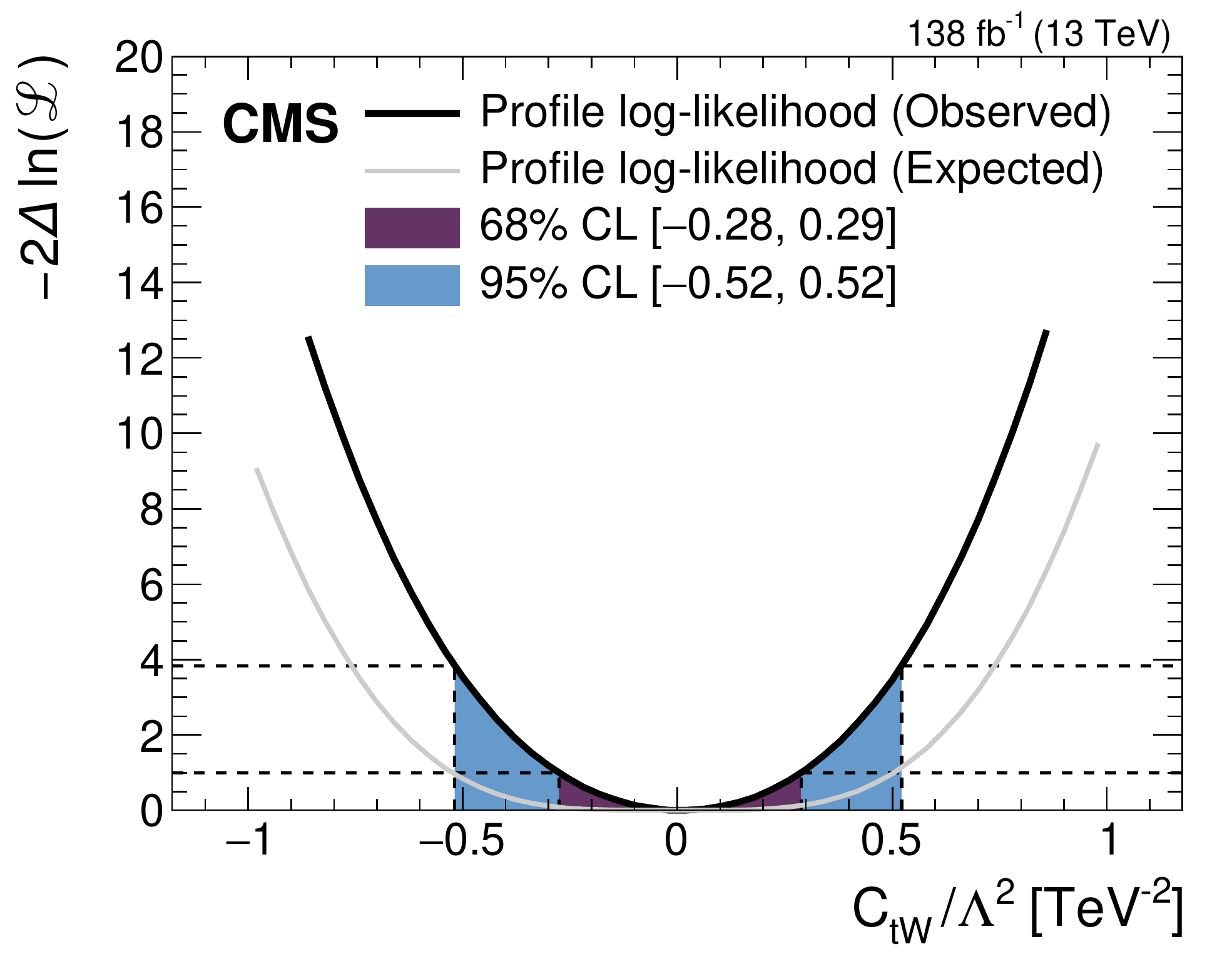} \\
  \includegraphics[width=0.49\textwidth]{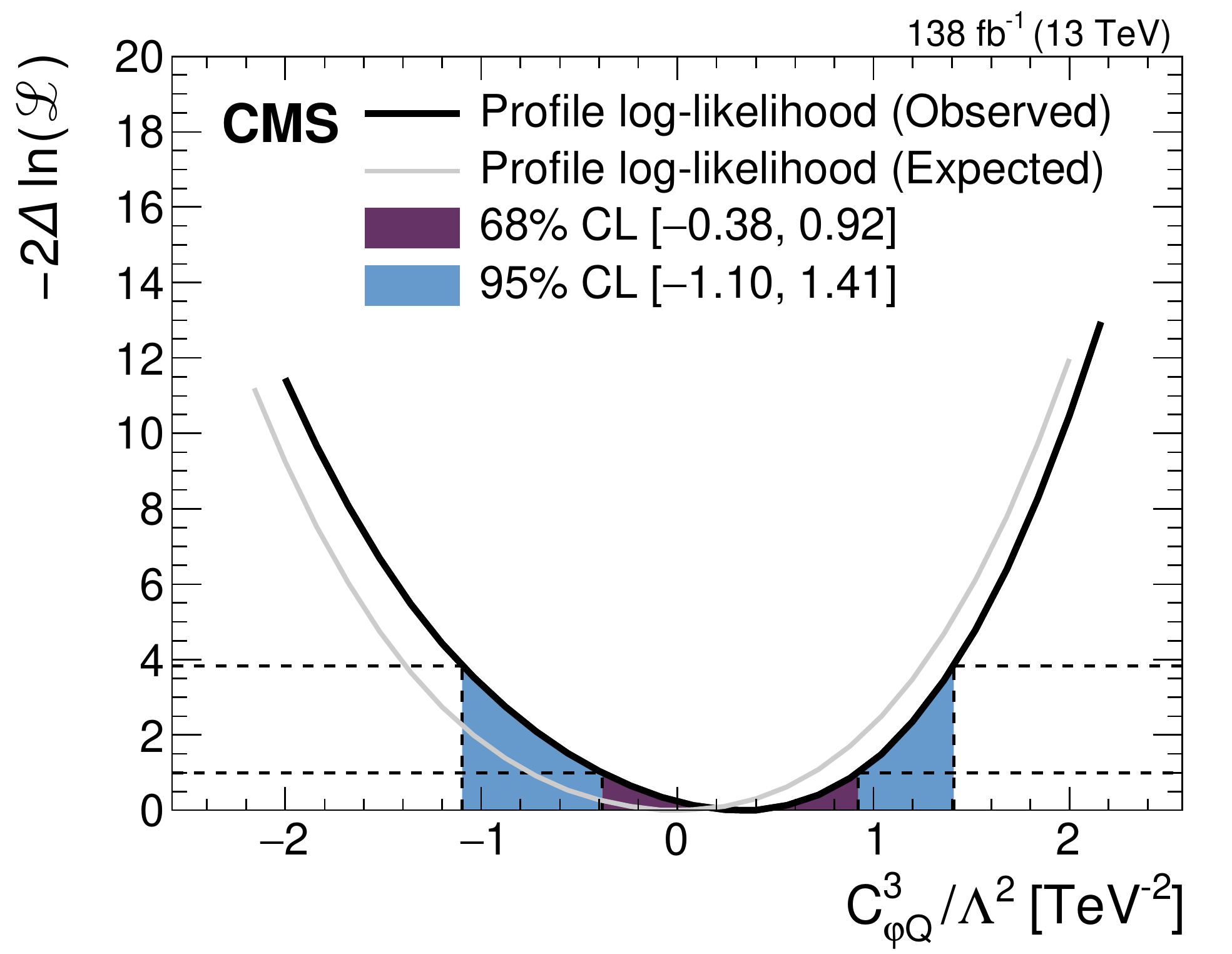} 
  \includegraphics[width=0.49\textwidth]{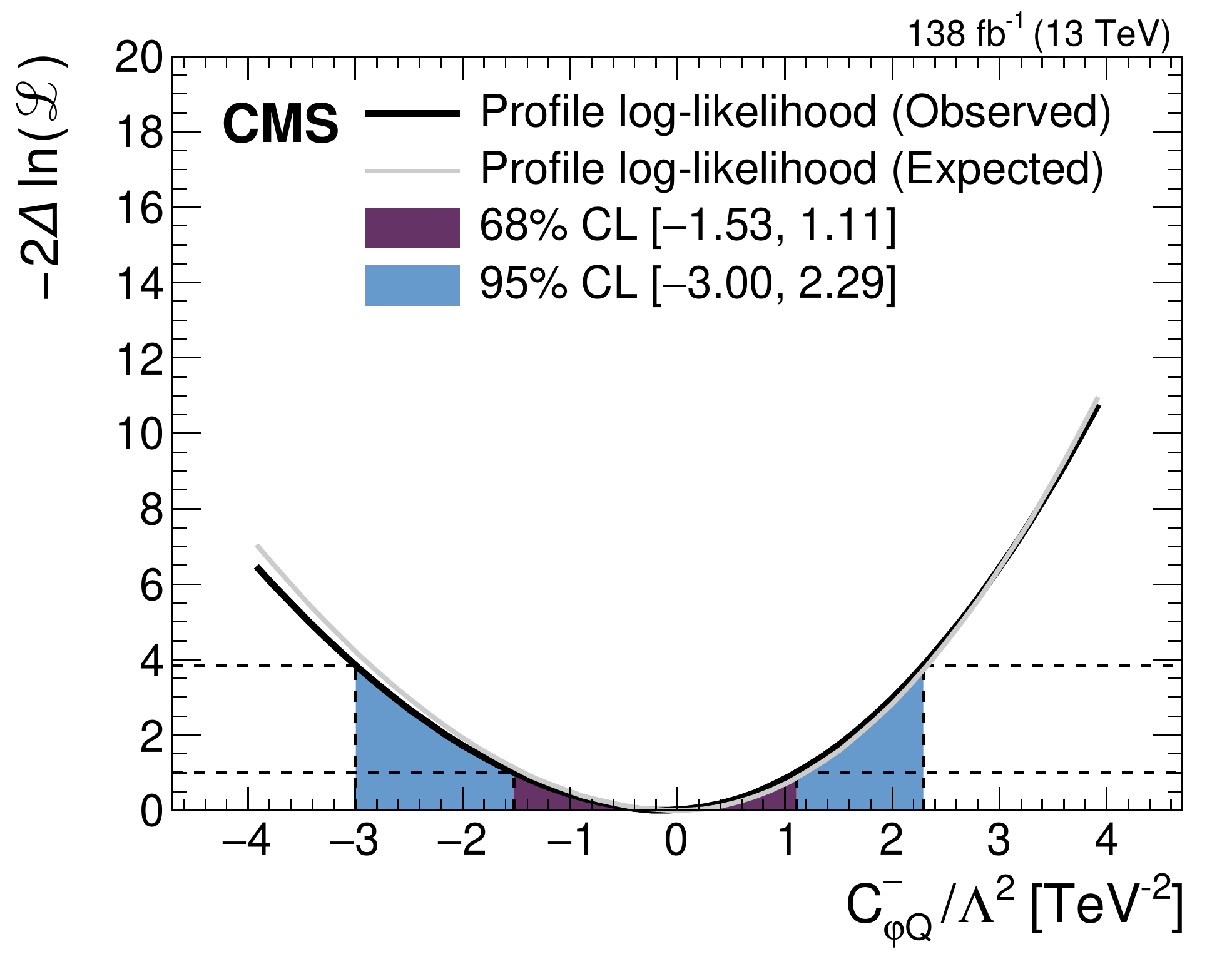} \\
  \includegraphics[width=0.49\textwidth]{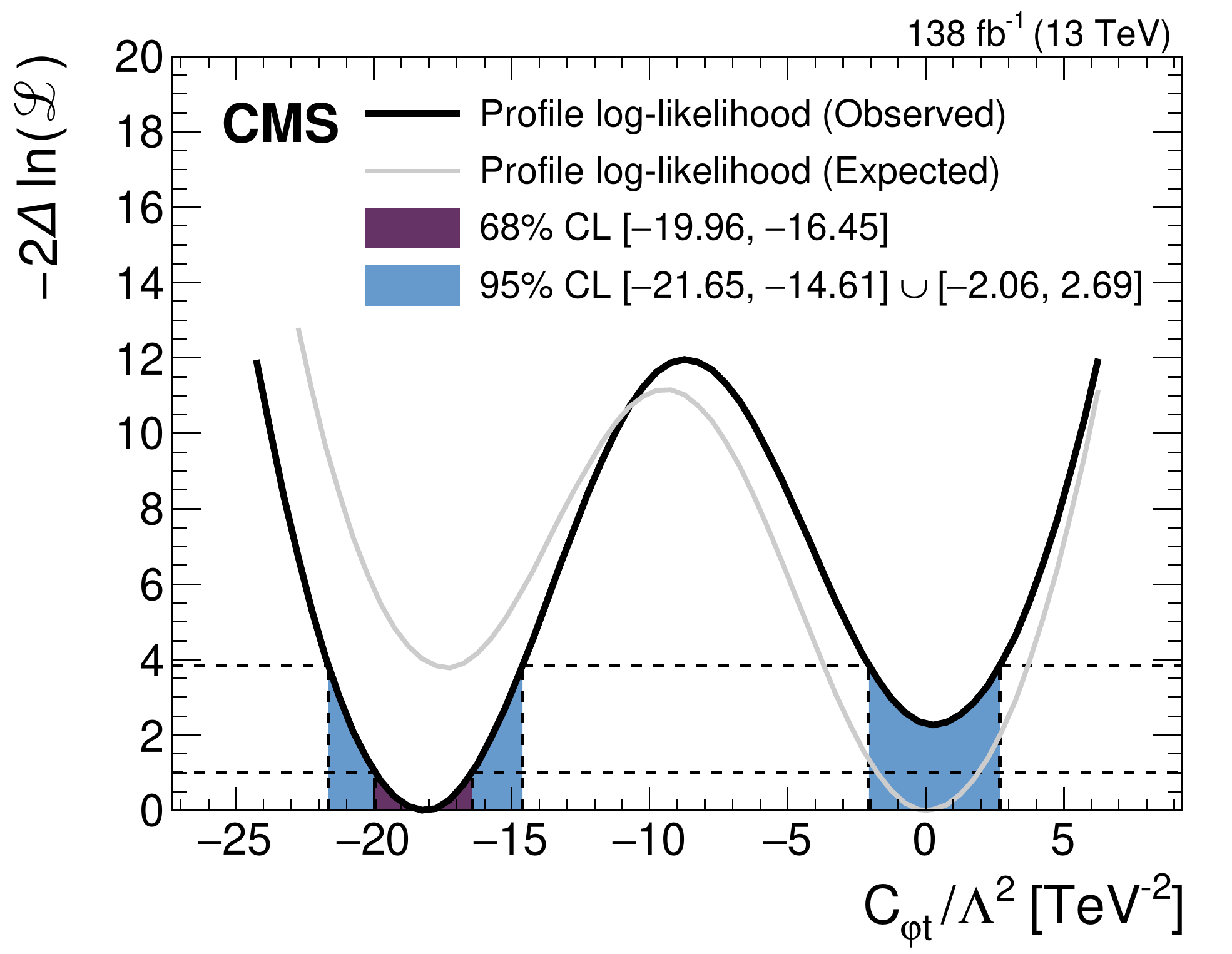} 
\caption{Observed (thick black lines) and expected (thin gray lines) one-dimensional scans of the negative log-likelihood as a function of each of the five WCs, while fixing the other WCs to their SM values of zero. The 68 and 95\% \CL confidence intervals are indicated by the colored areas.}
\label{fig:scan1D}
\end{figure}

\begin{figure}[htbp]
\centerline{
\includegraphics[width=0.49\textwidth]{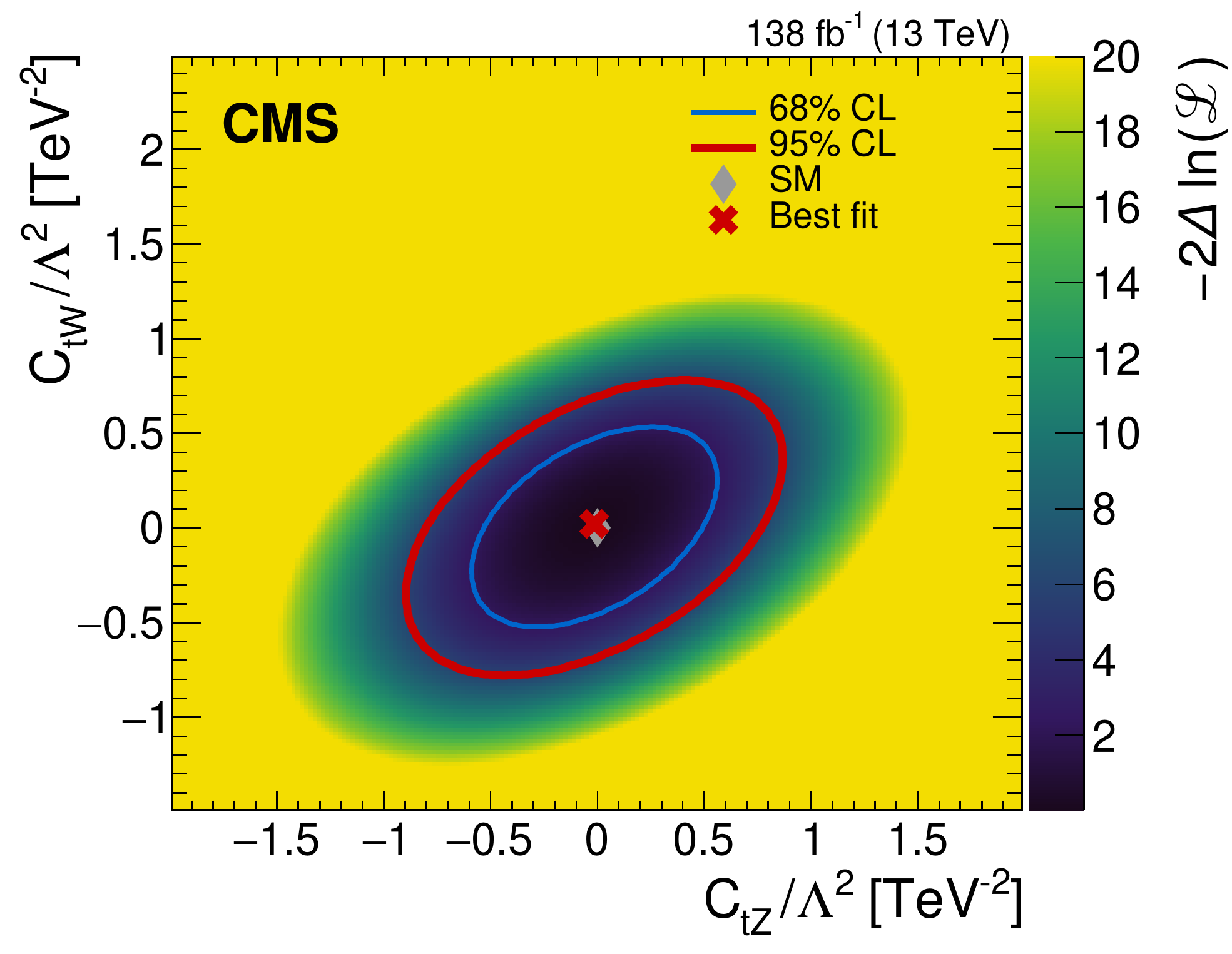}
\includegraphics[width=0.49\textwidth]{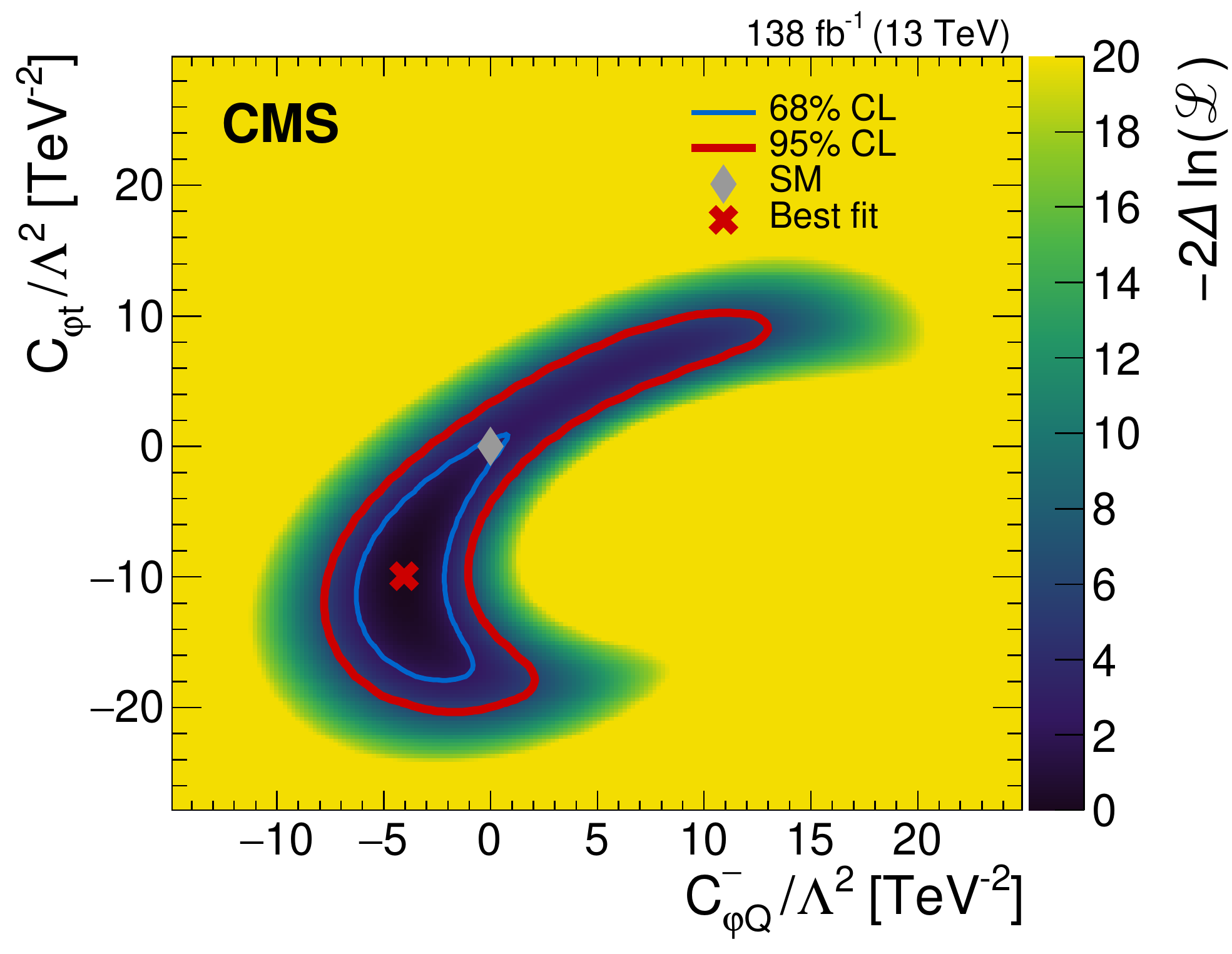}
}
\caption{Two-dimensional scans of the negative log-likelihood as a function of \ctz and \ctw (left), or as a function of \cpqm and \cpt (right), while fixing the other WCs to their SM values of zero. 
The SM and best fit points are indicated by diamond- and cross-shaped markers, respectively. 
The thin blue line and thick red line represent the 68 and 95\% \CL contours, respectively.}
\label{fig:scan2D}
\end{figure}

One-dimensional likelihood scans are shown for all WCs in Fig.~\ref{fig:scan1D}.
The 1D likelihood scan of \cpt exhibits a double-minima structure; while the data favor the negative minimum, they are compatible with the SM at 95\% CL. 
It was verified with MC pseudo-experiments generated under the SM hypothesis that the negative minimum is favored about 20\% of the time due to statistical fluctuations.
Two-dimensional likelihood scans are shown in Fig.~\ref{fig:scan2D} for the pairs of WCs that are most correlated.

\begin{table}[!hbtp]
  \centering
  \topcaption{
  \label{tab:intervals}
  Expected and observed 95\% \CL confidence intervals for all WCs. 
  The intervals in the first and second columns are obtained by scanning over a single WC, while fixing the other WCs to their SM values of zero. 
  The intervals in the third and fourth columns are obtained by performing a 5D fit in which all five WCs are treated as free parameters. 
  As explained in Section~\ref{sec:extraction}, the 1D intervals are obtained from separate fits to different observables in the \srtzq and \srttz, while the 5D intervals are obtained from a single fit.}
  \begin{tabular}{l{c}@{\hspace*{5pt}}ll{c}@{\hspace*{5pt}}ll}
    $\mathrm{WC} / \Lambda^2$&&\multicolumn{5}{c}{95\% \CL confidence intervals} \\
    $[\TeV^{-2}]$&& \multicolumn{2}{l}{Other WCs fixed to SM} && \multicolumn{2}{l}{5D fit} \\ \cline{3-4} \cline{6-7}
    && Expected & Observed && Expected & Observed \\[\cmsTabSkip]
    \ctz    && $[-0.97,0.96]$ & $[-0.76,0.71]$ && $[-1.24,1.17]$ & $[-0.85,0.76]$\\
    \ctw    && $[-0.76,0.74]$ & $[-0.52,0.52]$ && $[-0.96,0.93]$ & $[-0.69,0.70]$\\
    \cpqt   && $[-1.39,1.25]$ & $[-1.10,1.41]$ && $[-1.91,1.36]$ & $[-1.26,1.43]$\\
    \cpqm   && $[-2.86,2.33]$ & $[-3.00,2.29]$ && $[-6.06,14.09]$ & $[-7.09,14.76]$\\
    \cpt    && $[-3.70,3.71]$ & $[-21.65,-14.61]\bigcup[-2.06,2.69]$ && $[-16.18,10.46]$ & $[-19.15,10.34]$\\
  \end{tabular}
\end{table}

\begin{table}[!hbtp]
\begin{center}
\caption{Impacts from different groups of sources of systematic uncertainty on each individual WC.
To estimate the impact of a given group, the corresponding sources of systematic uncertainty are excluded, the 1D fits to the data are repeated, and the reduction in the width of the confidence interval is quoted for each WC\@. The values are given in percent.}
\label{tab:impact_wcs}
\begin{tabular}{lrrrrr}
Source & \ctz & \ctw & \cpqt & \cpqm & \cpt\\
\hline \\[-2.3ex]
\tZq normalization & $<$0.1 & $<$0.1 & $1.2 $ & $0.1$ & $0.8$\\
\ttZ normalization & $0.6$ & $<$0.1 & $0.4$ & $37$ & $38$\\ 
\tWZ normalization & $0.1$ & $0.1$ & $<$0.1 & $0.7$ & $2.1$\\ 
Background normalizations & $<$0.1 & $<$0.1 & $6.9$ & $3.6$ & $6.8$\\ 
NPL background estimation & $1.4$ & $0.2$ & $5.6$ & $0.3$ & $3.8$\\ 
Jet energy scale & $<$0.1 & $<$0.1 & $0.8$ & $0.7$ & $2.3$\\ 
Jet energy resolution & $<$0.1 & $<$0.1 & $<$0.1 & $<$0.1 & $1.4$\\
\ptmiss & $<$0.1 & $<$0.1 & $<$0.1 & $<$0.1 & $0.2$\\ 
\cPqb\ tagging & $<$0.1 & $<$0.1 & $0.9$ & $2.0$ & $0.3$\\ 
Other (experimental) & $<$0.1 & $<$0.1 & $1.6$ & $0.8$ & $0.6$\\
Lepton identification and isolation & $0.4$ & $0.4$ & $1.2$ & $2.2$ & $0.8$\\ 
Theory & $2.1$ & $1.1$ & $0.4$ & $0.9$ & $0.9$\\
\end{tabular}
\end{center}
\end{table}

\clearpage

\section{Summary}
\label{sec:summary}
A search for new top quark interactions has been performed within the framework of an effective field theory (EFT) using the associated production of either one or two top quarks with a \PZ boson in multilepton final states.
The data sample corresponds to an integrated luminosity of \usedLumi of proton-proton collisions at $\sqrt{s} = 13\TeV$ collected by the CMS experiment. 
Five dimension-six operators modifying the electroweak interactions of the top quark were considered. 
The event yields and kinematic properties of the signal processes were parameterized with Wilson coefficients (WCs) describing the interaction strengths of these operators.

A multivariate analysis relying upon machine-learning techniques was designed to enhance the sensitivity to effects arising from the EFT operators.
A multiclass neural network was trained to distinguish between standard model (SM) processes and was used to define three subregions enriched in \tZq, \ttZ, and background events. 
Additional neural networks were trained to separate events generated according to the SM from events generated with nonzero WC values, and were used to construct optimal observables.
This is the first time that machine-learning techniques accounting for the interference between EFT operators and the SM amplitude have been used in an LHC analysis.

Results were extracted from a simultaneous fit to data in six event categories.
Two confidence intervals were determined for each WC, one keeping the other WCs fixed to zero and the other treating all five WCs as free parameters. 
Two-dimensional contours were produced for pairs of WCs to illustrate their correlations.
All results are consistent with the SM at 95\% confidence level.

\begin{acknowledgments}
  We congratulate our colleagues in the CERN accelerator departments for the excellent performance of the LHC and thank the technical and administrative staffs at CERN and at other CMS institutes for their contributions to the success of the CMS effort. In addition, we gratefully acknowledge the computing  centers and personnel of the Worldwide LHC Computing Grid and other centers for delivering so effectively the computing infrastructure essential to our analyses. Finally, we acknowledge the enduring support for the construction and operation of the LHC, the CMS detector, and the supporting computing infrastructure provided by the following funding agencies: BMBWF and FWF (Austria); FNRS and FWO (Belgium); CNPq, CAPES, FAPERJ, FAPERGS, and FAPESP (Brazil); MES (Bulgaria); CERN; CAS, MoST, and NSFC (China); MINCIENCIAS (Colombia); MSES and CSF (Croatia); RIF (Cyprus); SENESCYT (Ecuador); MoER, ERC PUT and ERDF (Estonia); Academy of Finland, MEC, and HIP (Finland); CEA and CNRS/IN2P3 (France); BMBF, DFG, and HGF (Germany); GSRT (Greece); NKFIA (Hungary); DAE and DST (India); IPM (Iran); SFI (Ireland); INFN (Italy); MSIP and NRF (Republic of Korea); MES (Latvia); LAS (Lithuania); MOE and UM (Malaysia); BUAP, CINVESTAV, CONACYT, LNS, SEP, and UASLP-FAI (Mexico); MOS (Montenegro); MBIE (New Zealand); PAEC (Pakistan); MSHE and NSC (Poland); FCT (Portugal); JINR (Dubna); MON, RosAtom, RAS, RFBR, and NRC KI (Russia); MESTD (Serbia); SEIDI, CPAN, PCTI, and FEDER (Spain); MOSTR (Sri Lanka); Swiss Funding Agencies (Switzerland); MST (Taipei); ThEPCenter, IPST, STAR, and NSTDA (Thailand); TUBITAK and TAEK (Turkey); NASU (Ukraine); STFC (United Kingdom); DOE and NSF (USA).
  
  \hyphenation{Rachada-pisek} Individuals have received support from the Marie-Curie program and the European Research Council and Horizon 2020 Grant, contract Nos.\ 675440, 724704, 752730, 758316, 765710, 824093, and COST Action CA16108 (European Union); the Leventis Foundation; the Alfred P.\ Sloan Foundation; the Alexander von Humboldt Foundation; the Belgian Federal Science Policy Office; the Fonds pour la Formation \`a la Recherche dans l'Industrie et dans l'Agriculture (FRIA-Belgium); the Agentschap voor Innovatie door Wetenschap en Technologie (IWT-Belgium); the F.R.S.-FNRS and FWO (Belgium) under the ``Excellence of Science -- EOS" -- be.h project n.\ 30820817; the Beijing Municipal Science \& Technology Commission, No. Z191100007219010; the Ministry of Education, Youth and Sports (MEYS) of the Czech Republic; the Deutsche Forschungsgemeinschaft (DFG), under Germany's Excellence Strategy -- EXC 2121 ``Quantum Universe" -- 390833306, and under project number 400140256 - GRK2497; the Lend\"ulet (``Momentum") Program and the J\'anos Bolyai Research Scholarship of the Hungarian Academy of Sciences, the New National Excellence Program \'UNKP, the NKFIA research grants 123842, 123959, 124845, 124850, 125105, 128713, 128786, and 129058 (Hungary); the Council of Science and Industrial Research, India; the Latvian Council of Science; the Ministry of Science and Higher Education and the National Science Center, contracts Opus 2014/15/B/ST2/03998 and 2015/19/B/ST2/02861 (Poland); the National Priorities Research Program by Qatar National Research Fund; the Ministry of Science and Higher Education, project no. 0723-2020-0041 (Russia); the Programa Estatal de Fomento de la Investigaci{\'o}n Cient{\'i}fica y T{\'e}cnica de Excelencia Mar\'{\i}a de Maeztu, grant MDM-2015-0509 and the Programa Severo Ochoa del Principado de Asturias; the Stavros Niarchos Foundation (Greece); the Rachadapisek Sompot Fund for Postdoctoral Fellowship, Chulalongkorn University and the Chulalongkorn Academic into Its 2nd Century Project Advancement Project (Thailand); the Kavli Foundation; the Nvidia Corporation; the SuperMicro Corporation; the Welch Foundation, contract C-1845; and the Weston Havens Foundation (USA). 
\end{acknowledgments}

\bibliography{auto_generated}
\cleardoublepage \appendix\section{The CMS Collaboration \label{app:collab}}\begin{sloppypar}\hyphenpenalty=5000\widowpenalty=500\clubpenalty=5000
\textbf{Deutsches Elektronen-Synchrotron, Hamburg, Germany}\\*[0pt]
N.~Tonon$^\dag$, H.~Aarup~Petersen, M.~Aldaya~Martin, P.~Asmuss, S.~Baxter, M.~Bayatmakou, O.~Behnke, A.~Berm\'{u}dez~Mart\'{i}nez, S.~Bhattacharya, A.A.~Bin~Anuar
% ...
\vskip\cmsinstskip
\dag: Deceased\\
% ...
\end{sloppypar}
\end{document}